\documentclass[useAMS,usenatbib]{mn2e}
\usepackage{natbib}
\usepackage{amsmath}
\usepackage{url}
\usepackage{longtable}
\usepackage{aas_macros}
\usepackage{amssymb}
\usepackage{graphicx}
\usepackage{deluxetable}

\newcommand{\about}{$\sim\!\!$~}
\newcommand{\kms}{\,km\,s$^{-1}$}

\def\lsim{\hbox{\rlap{\raise 0.425ex\hbox{$<$}}\lower 0.65ex\hbox{$\sim$}}}
\def\gsim{\hbox{\rlap{\raise 0.425ex\hbox{$>$}}\lower 0.65ex\hbox{$\sim$}}}

\def\arcsec{\hbox{$^{\prime\prime}$}}

\newcommand{\vsi}{\protect\hbox{$v_{\rm Si\,II}$}}

\newcommand{\vca}{\protect\hbox{$v_{\rm Ca\,H\&K}$}}
\newcommand{\vcaz}{\protect\hbox{$v_{\rm Ca\,H\&K}^{0}$}}

\newcommand\ion[2]{#1$\,${\small{#2}}\relax}

\title[SN~Ia Line Profiles]{On Spectral Line Profiles in Type I\lowercase{a} Supernova Spectra}

\author[Foley]{Ryan~J.~Foley$^{1}$\thanks{E-mail:rfoley@cfa.harvard.edu}\\
$^{1}$Harvard--Smithsonian Center for Astrophysics, 60 Garden Street, Cambridge, MA 02138, USA}

\begin{document}

\date{Accepted  . Received   ; in original form  }
\pagerange{\pageref{firstpage}--\pageref{lastpage}} \pubyear{2013}
\maketitle
\label{firstpage}

\begin{abstract}
  We present a detailed analysis of spectral line profiles in Type Ia
  supernova (SN~Ia) spectra.  We focus on the feature at \about 3500
  -- 4000~\AA, which is commonly thought to be caused by blueshifted
  absorption of Ca H\&K.  Unlike some other spectral features in SN~Ia
  spectra, this feature often has two overlapping (blue and red)
  components.  It is accepted that the red component comes from
  photospheric calcium.  However, it has been proposed that the blue
  component is caused by either high-velocity calcium (from either
  abundance or density enhancements above the photosphere of the SN)
  or \ion{Si}{II} $\lambda 3858$.  By looking at multiple data sets
  and model spectra, we conclude that the blue component of the Ca
  H\&K feature is caused by \ion{Si}{II} $\lambda 3858$ for most
  SNe~Ia.  The strength of the \ion{Si}{II} $\lambda 3858$ feature
  varies strongly with the light-curve shape of a SN.  As a result,
  the velocity measured from a single-Gaussian fit to the full line
  profile correlates with light-curve shape.  The velocity of the Ca
  H\&K component of the profile does not correlate with light-curve
  shape, contrary to previous claims.  We detail the pitfalls of
  assuming that the blue component of the Ca H\&K feature is caused by
  calcium, with implications for our understanding of SN~Ia
  progenitors, explosions, and cosmology.
\end{abstract}



\defcitealias{Maguire12}{M12}

\section{Introduction}\label{s:intro}

The spectral-energy distribution (SED) of a Type Ia supernova (SN~Ia)
near maximum brightness is relatively similar to that of a hot star.
An SN SED is predominantly a black body with line blanketing in the
ultraviolet.  There are also prominent spectral features associated
with absorption and emission from elements primarily generated in the
SN explosion.  These features typically have broad P-Cygni profiles,
although overlapping lines can produce larger and more complicated
profiles.

The exact SED of a SN~Ia depends on the velocity and density structure
of the SN ejecta \citep[e.g.,][]{Branch85}.  Since broad-band filters
sample portions of the SED, and measurements in such filters are used
to determine SN distances to ultimately measure cosmological
parameters \citep[e.g.,][]{Conley11, Suzuki12}, understanding SN~Ia
spectral features is important for precise cosmological measurements.

SN spectra also provide detailed information about the SN explosion,
progenitor composition, circumbinary environment, and reddening law
\citep[e.g.,][]{Hoflich98, Lentz01, Mazzali05:hvf, Tanaka08,
  Wang09:2pop, Foley12:csm, Hachinger12:10jn, Roepke12}.  Furthermore,
there is evidence that one can estimate the intrinsic colour of
SNe~Ia, and thus improve distance measurements through a better
estimate of the dust reddening by measuring the ejecta velocity of
SNe~Ia \citep{Foley11:vel}.  Ejecta velocity is measured from the
blueshifted position of spectral features.  For both cosmology and SN
physics, it is important to have a precise understanding of SN~Ia
SEDs.

At optical wavelengths, the two most prominent features in a
maximum-light spectrum of a SN~Ia are at \about 3750 and 6100~\AA,
respectively.  The latter is thought to be from \ion{Si}{II}
$\lambda\lambda 6347$, 6371 ($gf$-weighted rest wavelength of
6355~\AA), and is the hallmark spectral feature of a SN~Ia.  The
former, at rest-frame wavelengths of \about 3500 -- 4000~\AA, is
generally attributed to blueshifted absorption from Ca H\&K
($gf$-weighted rest wavelength of 3945~\AA).  However, the line
profile of this feature is complicated, often times displaying
shoulders, a flat bottom, a ``split'' profile, and/or two distinct
absorption components.  There is broad consensus that the red
component of the profile is from Ca H\&K at a ``photospheric''
velocity, i.e., a velocity similar to that of the ejecta at close to
the $\tau = 2/3$ surface, which is typically about 12000~\kms\ near
maximum light.  However, there is no clear consensus to the origin of
the blue component.  Previous studies have attributed the blue
absorption component to either ``high-velocity'' (HV) Ca H\&K
absorption \citep[\about 18,000~\kms; e.g.,][]{Hatano99:94d,
  Garavini04, Branch05, Branch07, Stanishev07:03du, Chornock08,
  Tanaka08, Tanaka11, Parrent12}, where the absorption comes from a
region at high velocity within the SN ejecta that has high-density
calcium, and to \ion{Si}{II} $\lambda\lambda 3854$, 3856, 2863
\citep[$gf$-weighted rest wavelength of 3858~\AA; e.g.,][]{Kirshner93,
  Hoflich95, Nugent97, Lentz00, Wang03, Altavilla07}.  Since calcium
and silicon produce the strongest features in SN~Ia spectra near
maximum brightness, both interpretations are worth investigation.  For
convenience, we will generally refer to this feature as the ``Ca H\&K
feature.''

There are several cases of clear HV material in SNe~Ia.  Observations
showing multiple components to the \ion{Si}{II} $\lambda 6355$ line
profile \citep[e.g.,][]{Mazzali05:99ee, Altavilla07, Garavini07:05cf,
  Stanishev07:03du, Wang09:05cf, Foley12:09ig} or strong and quickly
varying HV \ion{O}{I} $\lambda 7774$ \citep{Altavilla07, Nugent11} are
perhaps the cleanest way to detect HV material since there are no
other strong lines just blueward of \ion{Si}{II} $\lambda 6355$ and
\ion{O}{I} $\lambda 7774$.  Other detections have been made by
observing the Ca NIR triplet, often through spectropolarimetry
\citep[e.g.,][]{Hatano99:94d, Li01:00cx, Kasen03, Wang03, Gerardy04,
  Mazzali05:hvf}, however there are several subtleties to this
feature.

HV features must be caused by abundance and/or density enhancements in
layers of the ejecta above the SN photosphere.  Two distinct
``layers'' of material within a smooth density profile (i.e., an
abundance enhancement) would necessarily be caused by the explosion,
and observations of HV features could therefore restrict the possible
explosion models.  However, \citet{Mazzali05:99ee} suggested that
abundance differences alone cannot reproduce the strength of the HV
features, and therefore there must be a density enhancement.  Density
enhancements may be either caused by the explosion causing over-dense
blobs or shells of material or by sweeping up circumbinary material
\citep[e.g.,][]{Gerardy04, Mazzali05:hvf, Quimby06:05cg}.
Spectropolarimetric observations have indicated that HV Ca NIR triplet
features are probably caused from the explosion
\citep[e.g.,][]{Kasen03, Wang03, Chornock08}.  Because of its
wavelength, it is difficult to obtain high-quality spectropolarimetric
measurements of the Ca H\&K feature.  None the less, \citet{Wang03}
was able to make such a measurement, and the polarization spectrum
suggested that the blue component of the Ca H\&K feature was from
\ion{Si}{II} $\lambda 3858$ for SN~2001el.

Using the large CfA sample of SN~Ia spectra \citep{Blondin12},
\citet*{Foley11:vgrad} determined that the velocity of \ion{Si}{II}
$\lambda 6355$, \vsi, and the velocity of the red component of the Ca
H\&K feature, \vca, at maximum light correlated with intrinsic colour,
but did not correlate with light-curve shape (and thus luminosity).
However, they did not find statistically significant evidence of a
linear correlation between the pseudo-equivalent width of the Ca H\&K
feature and intrinsic colour.  Using SDSS--II Supernova Survey and
Supernova Legacy Survey data, \citet{Foley12:vel} confirmed these
trends with high-redshift SNe~Ia.  They also noted a slight
(2.4-$\sigma$ significant) trend between the maximum-light \vca\
(\vcaz) and host-galaxy mass.

\citet[hereafter \citetalias{Maguire12}]{Maguire12} presented a sample
of maximum-light low-redshift SN~Ia spectra obtained with the
\emph{Hubble Space Telescope} (\emph{HST}).  After various quality
cuts, the sample consisted of 16 spectra of 16 SNe~Ia.  These spectra
covered the Ca H\&K feature, but did not cover wavelengths near
\ion{Si}{II} $\lambda 6355$.  Unlike \citet{Foley11:vgrad} and
\citet{Foley12:vel}, which presumed that the red component of the Ca
H\&K feature was representative of photospheric calcium (and thus the
wavelength of the maximum absorption of this component represented
\vca), \citetalias{Maguire12} fit a single Gaussian to the entire
profile to measure \vca.  Among other claims, \citetalias{Maguire12}
reported a linear relationship between \vca\ and light-curve shape
(3.4-$\sigma$ significant).  Furthermore, they claimed that after
correcting for the relation between light-curve shape and \vca, there
is no correlation between \vca\ and host-galaxy mass.

In this paper, we examine the claims of \citetalias{Maguire12} with
particular scrutiny to the details of the Ca H\&K profile.  In
Section~\ref{s:cahk}, we re-examine the \citetalias{Maguire12} sample.
We confirm a difference in the Ca H\&K line profile for SNe~Ia with
different light-curve shapes, but show that the difference is
primarily in the blue component.  We also conclude that
single-Gaussian fits to the Ca H\&K feature give biased, unphysical
velocity measurements.  In Section~\ref{s:synow}, we provide simple
models of calcium and silicon features in a SN~Ia spectrum.  Trends in
the spectra indicate that the blue component of the Ca H\&K feature is
likely from \ion{Si}{II} $\lambda 3858$.  In Section~\ref{s:m12}, we
perform further analysis with the \citetalias{Maguire12} sample,
finding systematic biases in single-Gaussian velocity measurements
appear to be present in the \citetalias{Maguire12} analysis.  In
Section~\ref{s:cfa}, we re-examine the CfA spectral sample, providing
further evidence that (1) the blue component of the Ca H\&K feature is
from \ion{Si}{II} $\lambda 3858$ absorption and (2) there is no
evidence for a correlation between \vca\ and light-curve shape.
Finally, we examine the spectra of the well-observed SN~2011fe and
SN~2010ae, a very low-velocity SN~Iax, in Section~\ref{s:add}.
Although one cannot uniquely claim that the blue component of the Ca
H\&K profile is from \ion{Si}{II} $\lambda 3858$ for SN~2011fe, it
must be the the case for SN~2010ae.  We discuss implications of this
result and conclude in Section~\ref{s:conc}.


\section{The Ca H\&K  Line Profile}\label{s:cahk}

As noted above, the spectral feature at rest-frame wavelengths of
\about 3500 -- 4000~\AA\ in SN~Ia spectra often has structure such as
shoulders and multiple components.  The main absorption is thought to
be from Ca H\&K (from the photosphere and possibly from a HV
component) and \ion{Si}{II} $\lambda 3858$.  We will refer to this
feature as the Ca H\&K feature, although there may be additional
species contributing to it.

In this section, we will examine the Ca H\&K feature in detail.  To do
this, we use the \citetalias{Maguire12} spectra.  After testing their
claims of differences in the profile shape with light-curve shape, we
examine the different results one gets depending on the method of
fitting the line profile.

\begin{figure}
\begin{center}
\includegraphics[angle=90,width=3.2in]{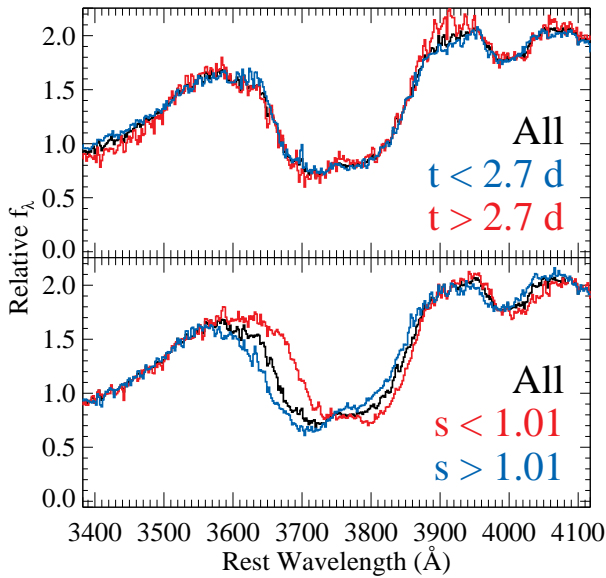}
\caption{Median spectra from the \citetalias{Maguire12} sample.  The
  black curve is the median spectrum from their full sample.  In the
  top panel, the blue and red curves represent the median spectra
  taken from early and late subsamples, respectively, while in the
  bottom panel, they represent low and high stretch (corresponding to
  low and high luminosity) subsamples, respectively.}\label{f:maguire}
\end{center}
\end{figure}

\citetalias{Maguire12} suggested that \vcaz\ depends on the light-cure
shape (and therefore peak luminosity) of the SN.  Using the WISERep
database \citep{Yaron12}, we obtained most of the spectra presented by
\citetalias{Maguire12}\footnote{The spectra of SNe~2011by and 2011fe
  were not included in the database, and are therefore excluded from
  our analysis.  But we do not expect including them in our analysis
  would change our results much.}.  We exclude all SNe that
\citetalias{Maguire12} do not use in their final analysis, including
PTF10ufj, which only has a redshift determined by SN spectral feature
matching.  In total, there are 14 spectra of 14 SNe~Ia in the final
\citetalias{Maguire12} sample.  In Figure~\ref{f:maguire}, we present
median spectra from the \citetalias{Maguire12} sample.  The Ca H\&K
feature has two clear minima (at \about 3720 and 3800~\AA,
respectively) in the median spectrum.  These data appear to be an
excellent sample for studying the Ca H\&K profile shape.

We also generated median spectra for subsets of the full sample.
First, we split the sample by phase.  Since the velocity of SN
features typically decreases monotonically with time because of the
receding photosphere, one expects lower velocity features at later
times.  The phase-split median spectra do not appear to be
significantly different from each other or the median spectrum from
the full sample.  This is likely the result of the
\citetalias{Maguire12} sample having a very narrow phase range.

We also split the sample by light-curve shape.  We split the sample by
$s = 1.01$ to match what was done by \citetalias{Maguire12}.  Here, we
see the same result that \citetalias{Maguire12} found and shows in
their Figures 5 and 7.  Namely, the low-stretch (corresponding to
faster-declines and lower luminosity) SNe have narrower, seemingly
lower velocity features than those of high-stretch SNe.

\begin{figure*}
\begin{center}
\includegraphics[angle=90,width=6.8in]{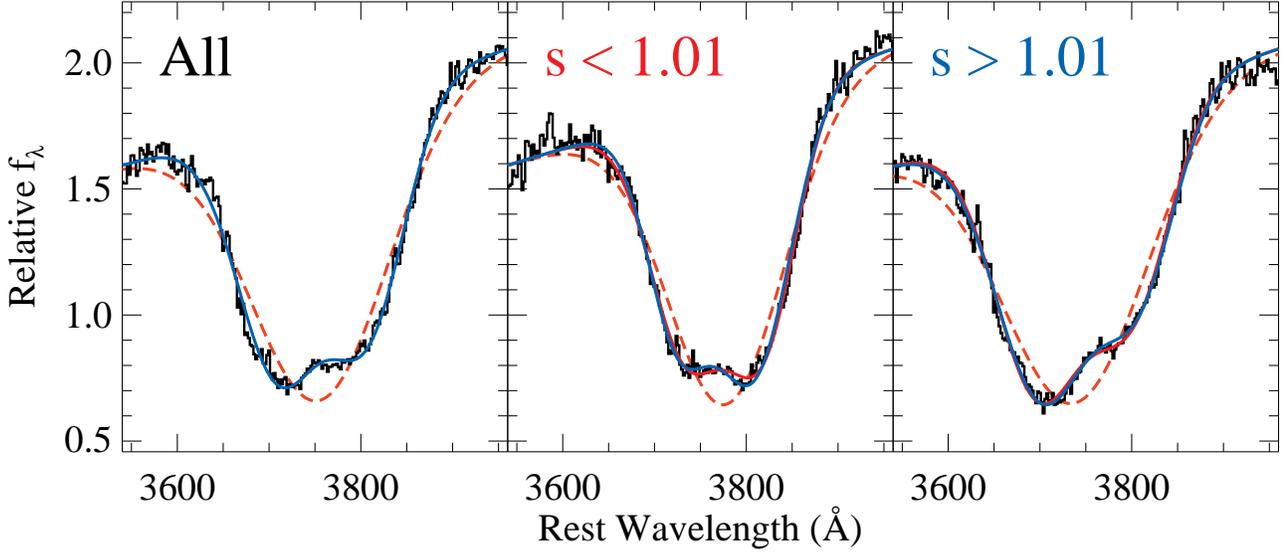}
\caption{Median spectra from the \citetalias{Maguire12} sample.  The
  left, center, and right panels show the median spectra from the full
  sample, the low-stretch subsample, and the high-stretch subsample,
  respectively.  The blue and red curves are double-Gaussian fits to
  the data.  The blue curve in the left panel and the red curves in
  the middle and right panels are the best fits with all parameters
  allowed to vary.  The blue curves in the middle and right panels
  represent fits where the centroid and width of the redder Gaussian
  were fixed to the best-fitting values for the full sample.  The
  orange dashed curves are single-Gaussian fits.}\label{f:dgauss}
\end{center}
\end{figure*}

Given the above difference, it is worth a detailed look at the line
profiles.  Despite coming from P-Cygni profiles, the line profiles
appear to be similar to the sum of two Gaussians, and performing such
a fit resulted in excellent matches to the profiles.  The absorption
component of a P-Cygni profile is very similar to a Gaussian, so using
Gaussians to fit the absorption is a reasonable choice.  In
Figure~\ref{f:dgauss}, we display the median spectra for the full
sample and the low/high-stretch subsamples.  We also display the
best-fitting double-Gaussian fits (after removing a linear
pseudo-continuum) to each line profile.  For each case, we performed a
six--parameter fit, allowing the centroid, width, and height of each
Gaussian to vary.  The centroid of each Gaussian corresponds roughly
to the characteristic velocity of that component.  Similarly, the
width of each feature corresponds to the velocity-width of the
absorbing region for that feature.  Finally, the height of each
feature is roughly related to the amount of absorbing material at a
given velocity.  The six--parameter double-Gaussian fits to each
profile are represented by the blue lines in Figure~\ref{f:dgauss}.

We also fit the low/high-stretch subsamples with two parameters fixed
and four allowed to vary.  The centroid and width (the parameters
related to velocity) of the redder Gaussian was fixed to match the
best-fitting values for the full-sample median spectrum, and the
remaining parameters (all parameters for the bluer Gaussian and the
height of the redder Gaussian) were allowed to vary.  These fits are
represented by the red lines in Figure~\ref{f:dgauss}.  Visually, the
six--parameter fit is not a significantly better representation of the
data than the four--parameter fit.  The reduced $\chi^{2}$ decreases
by 0.10 and 0.06 when changing from the six--parameter to the
four--parameter fit for the low and high-stretch subsamples,
respectively.  That is, the four--parameter fit has a smaller reduced
$\chi^{2}$ than the six--parameter fit (although only marginally
smaller), and thus, the subsamples and the full sample are completely
consistent with all having the same velocity for the red component.

\citetalias{Maguire12} argued that the difference in the red edge of
the Ca H\&K line profile was evidence that the subsamples have
different ejecta velocities.  But we have shown that simply varying
the height of the redder Gaussian (and the bluer Gaussian) are
sufficient to produce the red edge of the profile.  That is, the
apparent different in the red edge can be explained by different line
strengths rather than different line velocities, and thus a difference
in the red edge is not sufficient to distinguish different velocity
features.

We also attempted to fix the parameters of the bluer Gaussian, but
that did not result in good fits.  From these tests, we see that (1)
the red component does not necessarily have a different centroid (and
thus velocity) for the two subsamples and (2) the blue component does
have a different centroid.

We now turn to the difficulty of reducing these profiles to a single
parameter, namely velocity.  There have been two approaches to measure
velocities.  The first fits a single Gaussian to a line profile and
ascribes the centroid of the Gaussian to the velocity of the feature.
This method is used by many studies, including \citetalias{Maguire12}.
The alternative is to measure the wavelength of maximum absorption
(usually after some smoothing) to represent the velocity of the
feature.  This is the method described by \citet{Blondin06} and used
by \citet{Foley11:vgrad} and \citet{Foley12:vel}.  Although there are
many arguments to use either method, we will focus on the potential
systematic errors of using these methods when a feature has multiple
components like the Ca H\&K feature.

In Figure~\ref{f:dgauss}, we also show a single-Gaussian fit to the Ca
H\&K feature.  Besides being a poor representation of the data, the
centroid of the Gaussian is consistently intermediate to the two
components.  Usually, one wants to measure the photospheric velocity
for a given feature.  With that goal, the single Gaussian clearly
fails.  A single Gaussian, by measuring something intermediate to the
two components, measures nothing physical.  Furthermore, the centroid
of the single Gaussian is significantly affected by the blue
component.  The single Gaussian fits for the subsamples indicate that
the low-stretch SNe have significantly lower velocities than the
high-stretch SNe.  However, the double-Gaussian fits show that this is
not the case for the photospheric component.

\begin{figure}
\begin{center}
\includegraphics[angle=90,width=3.2in]{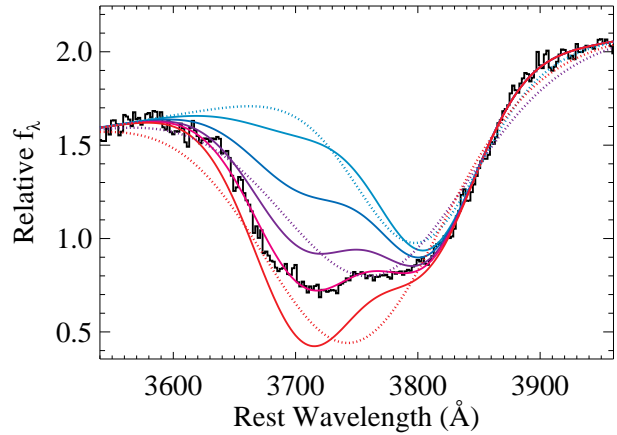}
\caption{Ca H\&K line profiles.  The black curve is the median
  spectrum from the full \citetalias{Maguire12} sample.  The solid
  lines are artificial line profiles created from two Gaussians where
  only the height of the blue component varies.  The dotted lines are
  single-Gaussian fits to the artificial profiles.}\label{f:gauss}
\end{center}
\end{figure}

To investigate the importance of the blue component to the measured
\vca\ from these two methods, we created artificial, but realistic,
line profiles.  In Figure~\ref{f:gauss}, we again show the median
spectrum from the \citetalias{Maguire12} sample.  We created a
double-Gaussian line profile to mimic the profile of the median
spectrum.  We then varied the height of the bluer Gaussian, but left
all other parameters fixed.  We display several example line profiles
in Figure~\ref{f:gauss}.  Visually, all of these line profiles appear
physically possible and represented in nature.  The full sample of
line profiles vary from having no blue component to having a blue
component that is about twice as strong as the red component.

\begin{figure}
\begin{center}
\includegraphics[angle=90,width=3.2in]{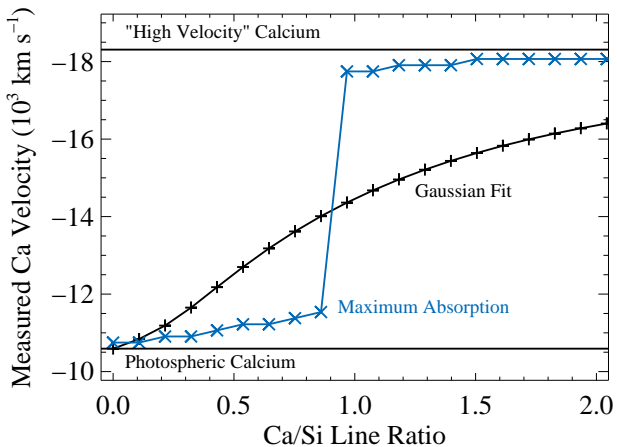}
\caption{Measured velocity for artificial line profiles.  The
  horizontal black lines represent the velocity of the two components
  (as measured from their centroid and assuming a rest wavelength of
  3945~\AA), with the lower and higher velocity components labelled
  ``Photospheric Calcium'' and ```High Velocity' Calcium,''
  respectively.  The black crosses represent the measured \vca\ from a
  single Gaussian to fit the profiles.  The blue X's represent the
  measured \vca\ from the wavelength of maximum
  absorption.}\label{f:vel_off}
\end{center}
\end{figure}

We fit single Gaussians to all artificial line profiles.  We display a
subset of these fits in Figure~\ref{f:gauss} (those that match the
subset of profiles displayed).  As expected, the stronger the blue
component, the bluer the centroid of the Gaussian.  In
Figure~\ref{f:vel_off}, we show the measured \vca\ from these Gaussian
fits.  Over the range we explore (from no blue component to a blue
component that is twice as strong as the red component), the measured
\vca\ changes by more than 5000~\kms.  Even when the blue component is
about a third as strong as the red component, the measured \vca\ is
\about 1000~\kms\ different from the true \vca.

We also measured the wavelength of maximum absorption.  This
wavelength is associated with the blue component when it is stronger
and quickly transition its association to the red component as the
blue component becomes weaker.  The measured \vca\ for our artificial
line profiles is shown in Figure~\ref{f:vel_off}.  Although this
method fails dramatically for strong blue components, the measured
\vca\ is relatively constant for line ratios less than one, with all
measured velocities $<$1000~\kms\ for all such cases.  There is a
slight bias (\about 150~\kms) for these measurements, some of which
can be explained by increasing flux of the pseudo-continuum with
wavelength.  Correcting for the pseudo-continuum removes much of the
bias, with the remaining bias related to the strength of the blue
component.

For cases where the red component is stronger than the blue
component, measuring the wavelength of maximum absorption is
significantly better at measuring the photospheric velocity than using
a Gaussian fit to the full profile.  In this regime, the wavelength of
maximum absorption is only minimally affected by the strength of the
blue component, while the Gaussian fit is significantly affected.  In
the regime of having a stronger blue component, the wavelength of
maximum absorption fails.  However, in this regime, the Gaussian fit
also fails, producing unphysical and significantly biased results.

Using the \citet{Foley11:vgrad} method of culling \vca\ measurements
that are not representative of the photospheric velocity, one should
have reliable \vca\ measurements, but will necessarily have an
incomplete sample.  A potential way to avoid this bias would be to
perform a double-Gaussian fit.  We have not investigated how this
method performs with noisy data.


\section{SYNOW Models}\label{s:synow}

To further understand the nature of the Ca H\&K feature, we use the SN
spectrum-synthesis code {\small SYNOW} \citep{Fisher97} to create
simple {\small SYNOW} spectral models.  We specifically use these
models to test how temperature can affect the profile and look for
trends between the Ca H\&K profile shape and other spectral features.
Although {\small SYNOW} has a simple, parametric approach to creating
synthetic spectra, it can provide insight on basic trends in SN SEDs.
To generate a synthetic spectrum, one inputs a blackbody temperature
($T_{\rm BB}$), a photospheric velocity ($v_{\rm ph}$), and for each
involved ion, an optical depth at a reference line, an excitation
temperature ($T_{\rm exc}$), the maximum velocity of the opacity
distribution ($v_{\rm max}$), and a velocity scale ($v_{e}$).  This
last variable assumes that the optical depth declines exponentially
for velocities above $v_{\rm ph}$ with an $e$-folding scale of
$v_{e}$.  The strengths of the lines for each ion are determined by
oscillator strengths and the approximation of a Boltzmann distribution
of the lower-level populations with a temperature of $T_{\rm exc}$.

\begin{figure*}
\begin{center}
\includegraphics[angle=90,width=6.8in]{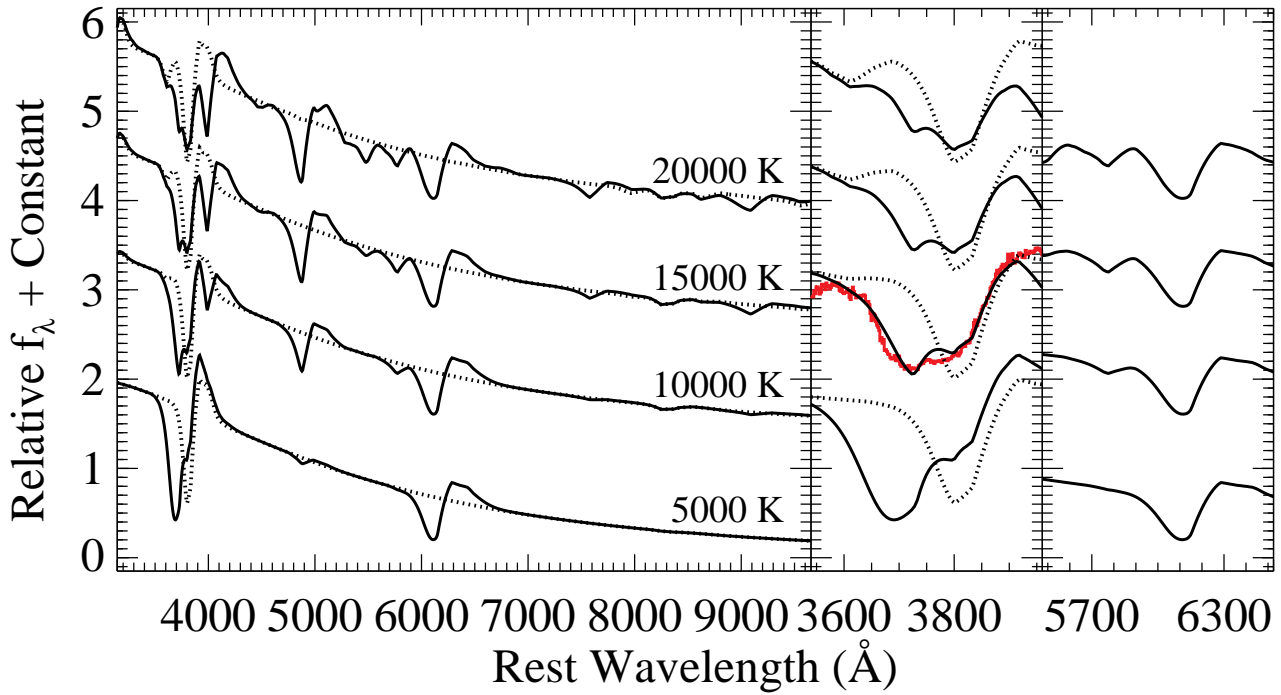}
\caption{{\small SYNOW} model spectra.  The dashed and solid curves
  represent models including only \ion{Ca}{II} and both \ion{Si}{II}
  and \ion{Ca}{II}, respectively.  The models only vary in their
  excitation temperature, which is labelled.  The middle and left
  panels show detailed views of the Ca H\&K feature and the redder
  \ion{Si}{II} complex, respectively.  The median spectrum from the
  \citetalias{Maguire12} sample is shown in the middle panel to
  demonstrate that the $T_{\rm exc} = 10$,000~K model has a similar Ca
  H\&K profile shape.}\label{f:sspec}
\end{center}
\end{figure*}

We produced models consisting of only \ion{Ca}{II} and with only
\ion{Si}{II} and \ion{Ca}{II} to isolate their affect on the profile
of the Ca H\&K feature.  For all models, we set $T_{\rm BB} =
10000$~K, $v_{\rm ph} = 10000$~\kms, $v_{\rm max} = 80000$~\kms, and
$v_{e} = 3000$ (for \ion{Si}{II}) and 2000~\kms\ (for \ion{Ca}{II}).
We chose $\tau = 5$ and 4 for \ion{Si}{II} and \ion{Ca}{II},
respectively.  These parameters were chosen such that when $T_{\rm
  exc} = 10000$~K, the model Ca H\&K line profile was visually similar
to that of the median spectrum of the \citetalias{Maguire12} sample.
Keeping all other parameters fixed, we varied $T_{\rm exc}$ from 5000
to 20000~K.  A subset of the models spanning this range are presented
in Figure~\ref{f:sspec}.

As seen in Figure~\ref{f:sspec}, the inclusion of \ion{Si}{II}
dramatically changes the Ca H\&K profile shape, making it stronger,
broader, and bluer.  Although the \ion{Si}{II} $\lambda 3858$ feature
may be stronger in the models than in real SN spectra, the \ion{Si}{II}
$\lambda 6355$ and the Ca H\&K features appear to have reasonable
strengths.

There is a clear spectral progression as the temperature changes.  We
note that for {\small SYNOW}, $T_{\rm bb}$ only changes the continuum
shape of the models and does not affect the strength of features.
Since {\small SYNOW} uses Ca H\&K as the reference calcium line, the
strength of the Ca H\&K absorption by definition does not change much
with $T_{\rm exc}$, and the entire calcium spectrum does not change
much over the temperatures probed.  Meanwhile the \ion{Si}{II}
spectrum changes significantly with varying $T_{\rm exc}$.  The
strength of the Ca H\&K absorption within the Ca H\&K feature (i.e.,
the strength of the red component) does change slightly with $T_{\rm
  exc}$ because of the strength of the \ion{Si}{II} $\lambda 3858$
emission changing the apparent Ca H\&K absorption.

In the red, there is the expected change in the ratio of the
\ion{Si}{II} $\lambda 5972$ and $\lambda 6355$ lines.  This ratio,
$\mathcal{R}$(Si), is highly correlated with luminosity and
light-curve shape \citep{Nugent95}.  As the \ion{Si}{II} $\lambda
5972$ feature becomes stronger, the \ion{Si}{II} $\lambda 3858$
feature becomes weaker.  For the {\small SYNOW} models,
$\mathcal{R}$(Si) increases with increasing $T_{\rm exc}$, while for
SNe~Ia, $\mathcal{R}$(Si) increases with decreasing $T$; this has been
previously noted \citep[e.g.,][]{Bongard08}, and is likely the result
of not simultaneously changing the opacity with $T_{\rm exc}$ and/or
non-local thermodynamic equilibrium effects.  However, the
\ion{Ca}{II} spectrum does not change significantly with $T_{\rm exc}$
and other model spectra show the same relation between \ion{Si}{II}
$\lambda 3858$ and \ion{Si}{II} $\lambda 5972$
\citep[e.g.,][]{Kasen07:asym, Blondin13}.  We therefore consider the
qualitative changes in the spectra to be correct, although the
corresponding temperatures may not be.  All models show that the
strength of \ion{Si}{II} $\lambda 3858$ and \ion{Si}{II} $\lambda
5972$ are anti-correlated; we will use this relation as the primary
model prediction.  We will later use $\mathcal{R}$(Si) as a proxy for
light-curve shape.

The excitation energy for the various \ion{Si}{II} lines also explain
the correlations between the various \ion{Si}{II} features.  The
\ion{Si}{II} $\lambda 3858$, \ion{Si}{II} $\lambda 5972$, and
\ion{Si}{II} $\lambda 6355$ features have excitation energies of 6.9,
10.0, and 8.1~eV, respectively.  Because the \ion{Si}{II} $\lambda
3858$ and \ion{Si}{II} $\lambda 5972$ features have very different
excitation energies and \ion{Si}{II} $\lambda 6355$ has an excitation
energy intermediate to the other two features, the strengths of the
\ion{Si}{II} $\lambda 3858$ and \ion{Si}{II} $\lambda 5972$ features
should change in opposite directions with changing temperature.  This
also explains the {\small SYNOW} results since {\small SYNOW} fixes
the strength of the reference feature, \ion{Si}{II} $\lambda 6355$.

In the middle and right-hand panels of Figure~\ref{f:sspec}, we show
the Ca H\&K feature and redder \ion{Si}{II} complex in detail.  Again,
it is clear that both $\mathcal{R}$(Si) and the strength of the
\ion{Si}{II} $\lambda 3858$ feature change in the way described above.

A SN photosphere is, of course, more complicated than the simple
{\small SYNOW} model.  Specifically, as the temperature changes over
the relevant range, the ionization of silicon (specifically the amount
of singly and doubly ionized silicon) changes.  Additionally, certain
features may be saturated (and possibly for only certain
temperatures).  Specifically, it is thought that \ion{Si}{II} $\lambda
6355$ is usually saturated.  However, it seems unlikely that the other
\ion{Si}{II} features are saturated.  Therefore, these additional
complexities should not change our interpretation that the strengths
of the \ion{Si}{II} $\lambda 3858$ and \ion{Si}{II} $\lambda 5972$
features are anti-correlated and change with temperature.

\begin{figure}
\begin{center}
\includegraphics[angle=90,width=3.2in]{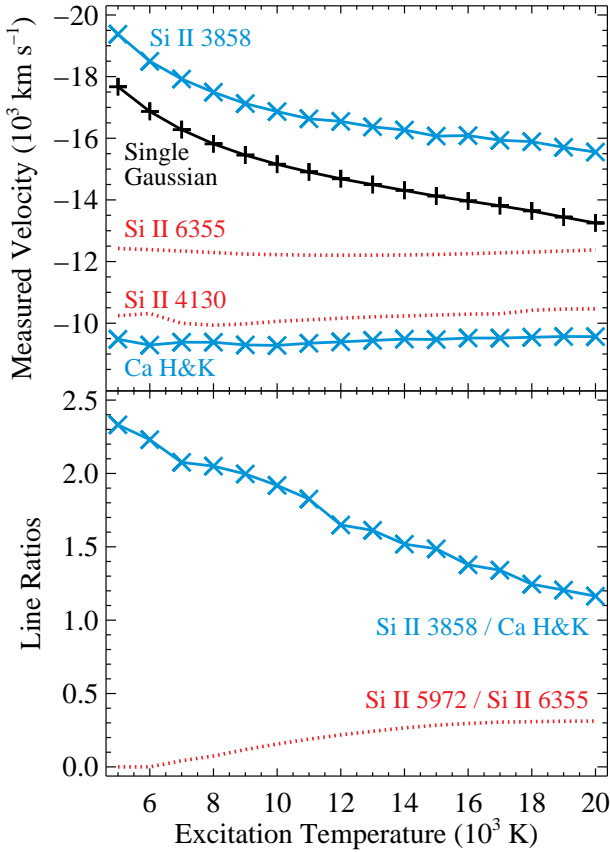}
\caption{\emph{Top Panel:} Measured velocity as a function of
  excitation temperature for {\small SYNOW} model spectra.  The red
  dotted lines represent the velocities measured with a single
  Gaussian for the \ion{Si}{II} $\lambda 4130$ and $\lambda 6355$
  lines.  The black crosses represent velocities measured with a
  single-Gaussian fit to the Ca H\&K feature.  The blue X's represent
  velocities measured with a double-Gaussian fit to the Ca H\&K
  feature, where the \ion{Si}{II} $\lambda 3858$ velocity is measured
  assuming its rest wavelength is 3945~\AA.  \emph{Bottom Panel:} The
  $\mathcal{R}$(Si) (red dotted line) and Si/Ca (blue X's) line ratios
  as a function of excitation temperature for the {\small SYNOW} model
  spectra.}\label{f:spar}
\end{center}
\end{figure}

We fit the \ion{Si}{II} $\lambda 4130$, $\lambda 5972$, and $\lambda
6355$ features in each model spectrum with single Gaussians.  Although
the line profiles are not exactly Gaussian, the fits are reasonable
approximations of the data, and the process is similar to what is done
in practice.  We also fit the Ca H\&K feature with both a single
Gaussian and a double Gaussian.  We show the measured velocity in
Figure~\ref{f:spar}.

The measured \ion{Si}{II} $\lambda 4130$, and $\lambda 6355$
velocities differ by at most 530 and 220~\kms\ over the entire
temperature range, respectively.  At the lowest $T_{\rm exc}$, the
\ion{Si}{II} $\lambda 5972$ feature is not strong enough to measure a
reliable velocity, but for the other temperatures, it differs by at
most 440~\kms.  These differences are encouraging since the
photospheric velocities did not change.

Fitting two Gaussians to the Ca H\&K feature, which should be better
at recovering the true velocity (see Section~\ref{s:cahk}), we see
that the red component, corresponding to Ca H\&K, has a measured
velocity range similarly small to that of the \ion{Si}{II} features
noted above.  Specifically, the maximum difference of measured Ca H\&K
velocities over all temperatures probed is only 290~\kms.  However,
the measured velocity for the \ion{Si}{II} $\lambda 3858$ feature
changes significantly with temperature.  Over the full temperature
range probed, the measured \ion{Si}{II} $\lambda 3858$ velocity ranges
from $-15$,550 to $-19,$380~\kms\ -- a difference of 3820~\kms, or
roughly an order of magnitude greater than that of the other features.

A single-Gaussian fit performs even worse.  Because of the changing
\ion{Si}{II} $\lambda 3858$ velocity and its varying strength, a
single-Gaussian fit to the Ca H\&K feature results in a \vca\ range of
$-13$,250 to $-17,670$~\kms\ over our chosen temperature range for a
maximum difference of 4420~\kms.  Since the \vca\ measured using a
single Gaussian can have dramatic differences even when there is no
change in physical velocities, this is even more reason to avoid this
technique.

Figure~\ref{f:spar} also shows the \ion{Si}{II} $\lambda 5972$ to
\ion{Si}{II} $\lambda 6355$ ($\mathcal{R}$(Si)) and \ion{Si}{II}
$\lambda 3858$ to Ca H\&K ratios, which we will call the Si/Ca ratio.
The range for $\mathcal{R}$(Si), from effectively zero (when
\ion{Si}{II} $\lambda 5972$ is difficult to discern) to \about 0.3, is
approximately the range seen by all SNe~Ia except SN~1991bg-like
objects \citep[e.g.,][]{Blondin12, Silverman12:lc}.  The Ca/Si ratio
has a range of 1.2 to 2.3.  The ratio is affected by both the strength
of the \ion{Si}{II} $\lambda 3858$ absorption and the \ion{Si}{II}
$\lambda 3858$ \emph{emission}, which fills in some of the Ca H\&K
absorption.  As noted above, the strength of the \ion{Si}{II} $\lambda
3858$ feature can have a large affect on the Ca H\&K line profile, and
even dominates for many temperatures.

\begin{figure}
\begin{center}
\includegraphics[angle=90,width=3.2in]{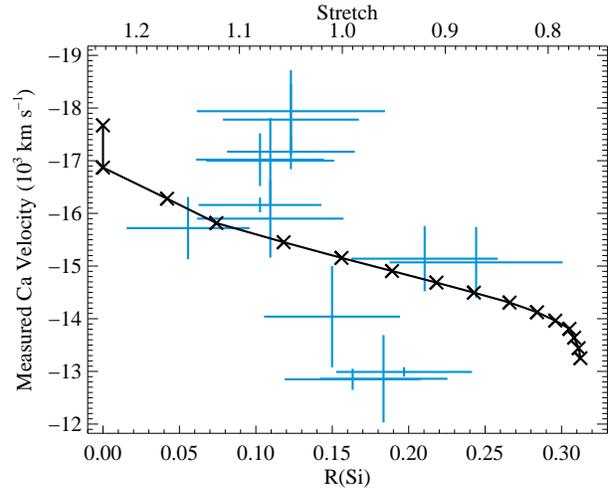}
\caption{Velocities measured with a single-Gaussian fit to the Ca H\&K
  feature as a function of $\mathcal{R}$(Si) for the {\small SYNOW}
  model spectra.  The \citetalias{Maguire12} data are also plotted in
  blue where the measured stretch is converted to $\mathcal{R}$(Si)
  with the top axis showing the scaling.}\label{f:si_cavel}
\end{center}
\end{figure}

Since \ion{Si}{II} $\lambda 5972$ and $\lambda 6355$ do not show
significant velocity differences with temperature, are relatively free
of contamination from other species, and $\mathcal{R}$(Si) is a good
indication of light-curve shape, we can use it as a proxy for
light-curve shape for our models.  Figure~\ref{f:si_cavel} shows \vca\
measured with a single Gaussian as a function of $\mathcal{R}$(Si).
The measured \vca\ decreases in amplitude with increasing
$\mathcal{R}$(Si), which corresponds to decreasing stretch and
luminosity.

Converting stretch to $\mathcal{R}$(Si), we can plot the
\citetalias{Maguire12} measurements on Figure~\ref{f:si_cavel}.  The
\citetalias{Maguire12} spectra do not cover the redder \ion{Si}{II}
features, so a direct measurement could not be made.  The
\citetalias{Maguire12} values, which use a single Gaussian to fit the
Ca H\&K feature, span a similar range of \vca\ and inferred
$\mathcal{R}$(Si) as the models, with the model trend going through
the middle of the data values.  The claimed trend of \vca\ with
light-curve shape is clear using the \citetalias{Maguire12} values.
However, this trend is similar to the trend generated by simply
changing the temperature in the {\small SYNOW} models.  We emphasize
that the true \vca\ is fixed for all models and the \vca\ measured
using a double-Gaussian fit varies only slightly over all models.  The
trend shown in Figure~\ref{f:si_cavel} is solely the result of the
method for measuring the velocity.  The underlying physical effect is
the changing strength of \ion{Si}{II} $\lambda 3858$ with temperature,
and thus light-curve shape.

There will undoubtedly be a true range of velocities in the data.  We
have not explored how the {\small SYNOW} models change when varying
several parameters, but it is clearly possible to reproduce the
\citetalias{Maguire12} trend through a combination of single Gaussian
fitting, an inherent velocity range, and relations between the Ca H\&K
profile shape with temperature and velocity.


\section{The M12 Sample}\label{s:m12}

With the insights of the above analysis, we re-analyse the individual
\citetalias{Maguire12} spectra.  In Section~\ref{s:cahk}, we showed
that fitting the Ca H\&K profile with a single Gaussian results in an
imprecise, biased, and unphysical measurement of the photospheric
velocity.  Instead, we fit the Ca H\&K profiles of the
\citetalias{Maguire12} spectra with two Gaussians.  As suggested
above, this method should provide relatively unbiased measurements of
the \ion{Si}{II} $\lambda 3858$ feature and the Ca H\&K feature.

We provide the best-fitting velocities for \ion{Si}{II} $\lambda 3858$
assuming that it is ``HV'' Ca H\&K and Ca H\&K in Table~\ref{t:data}.
We also provide the Si/Ca ratio in Table~\ref{t:data}.

\begin{figure}
\begin{center}
\includegraphics[angle=90,width=3.2in]{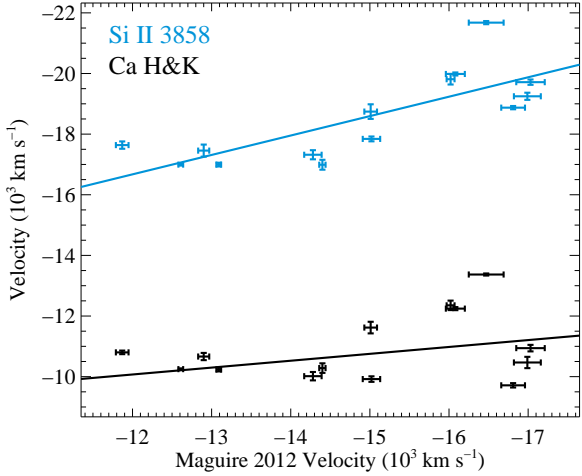}
\caption{\citetalias{Maguire12} Ca H\&K velocities measured using a
  single-Gaussian fit to the Ca H\&K feature compared to velocities
  measured with a double-Gaussian fit to the Ca H\&K feature for the
  \citetalias{Maguire12} sample.  The black and blue points represent
  the Ca H\&K and \ion{Si}{II} $\lambda 3858$ (assuming a rest
  wavelength of 3945~\AA) velocities from the double-Gaussian fit.
  The solid lines represent the best-fitting linear relationships for
  the data, corresponding to 1.5 and 3.1-$\sigma$ results for Ca H\&K
  and \ion{Si}{II} $\lambda 3858$, respectively.}\label{f:vel_vel}
\end{center}
\end{figure}

Figure~\ref{f:vel_vel} compares the velocities measured with the
double-Gaussian fit to those reported by \citetalias{Maguire12}.  The
\vca\ is systematically lower than the \citetalias{Maguire12} value
for the same SNe.  Similarly, the \ion{Si}{II} $\lambda 3858$ (when
treated as HV Ca H\&K) is systematically higher than the
\citetalias{Maguire12} value for the same SNe.  Performing a Bayesian
Monte-Carlo linear regression on the
\citetalias{Maguire12}--\ion{Si}{II} and \citetalias{Maguire12}--Ca
H\&K data sets \citep{Kelly07}, we find that 99.8 and 86.8 per cent of
the realizations have positive slopes, respectively, corresponding to
3.1-$\sigma$ and 1.5-$\sigma$ results, respectively.  That is, there
is significant evidence for a linear relation between the
\citetalias{Maguire12} \vca\ measurements and the \ion{Si}{II}
$\lambda 3858$ velocity, but no evidence for a linear relation between
the \citetalias{Maguire12} \vca\ measurements and the velocity of the
red component corresponding to absorption from photospheric calcium.

As expected from the results of Section~\ref{s:synow}, the
\citetalias{Maguire12} \vca\ measurements are intermediate to the
\ion{Si}{II} $\lambda 3858$ and Ca H\&K velocities, more closely track
the \ion{Si}{II} $\lambda 3858$ velocity than the Ca H\&K velocity, and
are systematically biased measurements of \vca.

For much of the analysis performed by \citetalias{Maguire12}, they
corrected their measured \vca\ to a maximum-light value, \vcaz, using
a single velocity gradient for all objects derived from the \vca\ and
effective phase measurements for their sample.  Using a large sample
where many objects had multiple spectra obtained near maximum light,
\citet{Foley11:vgrad} showed that \vcaz\ and the velocity gradient
were highly correlated with higher-velocity SNe also having
higher-velocity gradients.  Taking into account this relation, they
produced an equation which estimates \vcaz\ given \vca\ and phase.
Using the \citetalias{Maguire12} velocity gradient results in
differences between \vca\ and \vcaz\ that can be as large as
1150~\kms, and the median difference is 460~\kms.  However, if one
uses the \citet{Foley11:vgrad} relation to correct the velocity
measurements to have a common phase of 2.7~d (the median of the
sample), then the deviation between that value and \vca\ is at most
560~\kms, with a median absolute deviation of 120~\kms, both of which
are smaller than the typical uncertainty of 670~\kms\ for the
\citetalias{Maguire12} \vcaz\ measurements.  The combination of using
a single velocity gradient and extrapolating to maximum light (only
one spectrum in the sample has a phase before maximum brightness)
introduces unnecessary additional uncertainty.  Instead in all further
analysis, we use the raw \vca\ measurements, but add an additional
120~\kms\ uncertainty in quadrature to the reported uncertainty.

\begin{figure}
\begin{center}
\includegraphics[angle=90,width=3.2in]{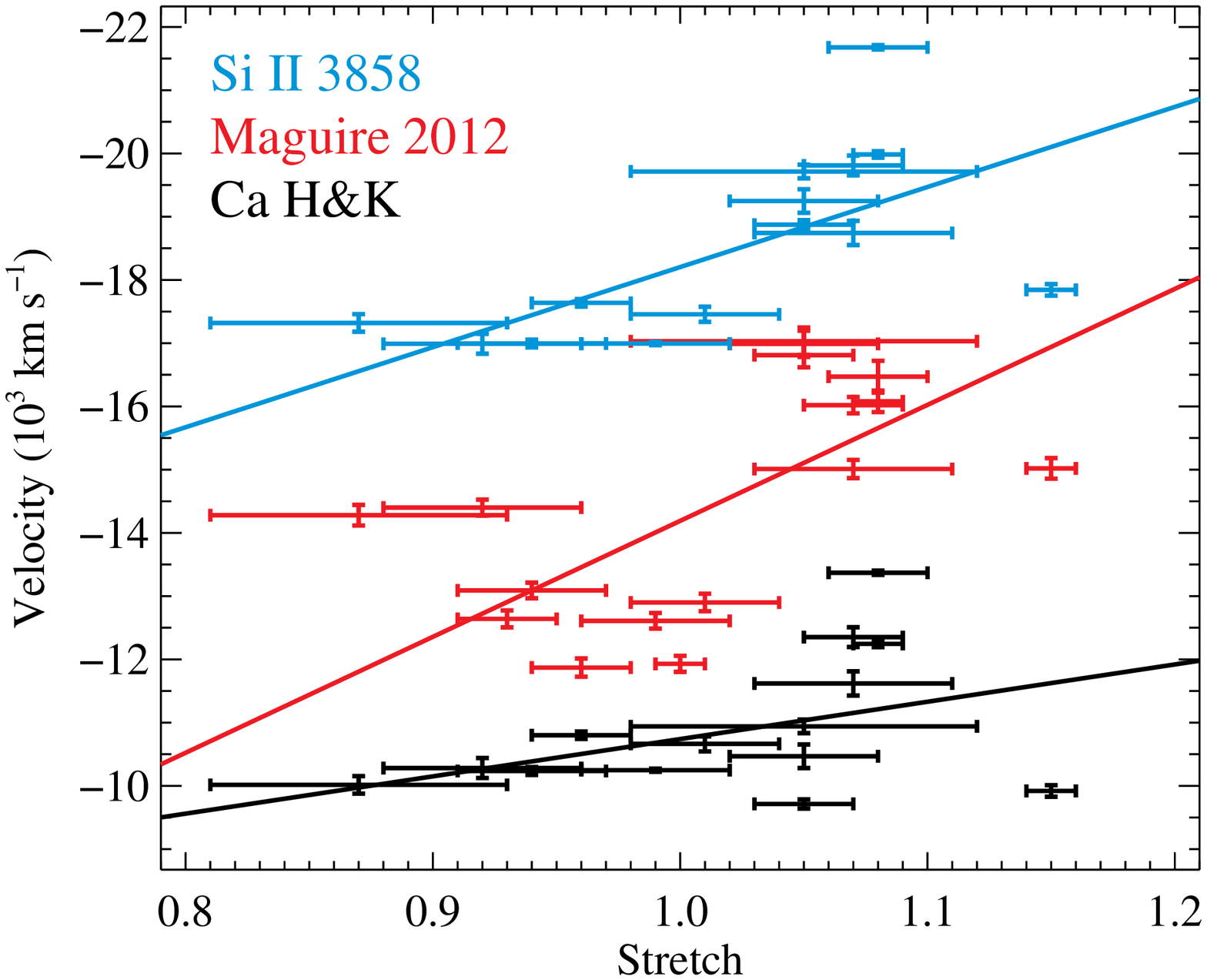}
\caption{Velocities measured with a single or double-Gaussian fit to
  the Ca H\&K feature as a function of stretch for the
  \citetalias{Maguire12} sample.  The single-Gaussian measurements are
  taken from \citetalias{Maguire12} and represented by the red data.
  The double-Gaussian measurements are shown as black and blue points
  for the Ca H\&K and \ion{Si}{II} $\lambda 3858$ (assuming a rest
  wavelength of 3945~\AA) components, respectively.  The solid lines
  represent the best-fitting linear relationships for the data,
  corresponding to 2.3, 2.3, and 1.4-$\sigma$ results for the
  \citetalias{Maguire12}, \ion{Si}{II} $\lambda 3858$, and Ca H\&K
  velocities, respectively.}\label{f:s_vel}
\end{center}
\end{figure}

Using our measurements of \vca, we can re-examine the
\citetalias{Maguire12} claim that \vca\ is correlated with light-curve
shape.  In Figure~\ref{f:s_vel}, we show the \ion{Si}{II} $\lambda
3858$, Ca H\&K, and \citetalias{Maguire12} velocity measurements as a
function of stretch.  This figure is similar to
Figure~\ref{f:si_cavel}.

Performing a Bayesian Monte--Carlo linear regression on the
\citetalias{Maguire12}, the \ion{Si}{II} $\lambda 3858$, and Ca H\&K
velocity measurements, we find that 97.7, 97.8, and 87.6 per cent of the
realizations have positive slopes, respectively, corresponding to
2.3-$\sigma$, 2.3-$\sigma$, and 1.5-$\sigma$ results, respectively.
We therefore find mild evidence that the \citetalias{Maguire12} and
\ion{Si}{II} $\lambda 3858$ velocity measurements are linearly related
to stretch.  We find no statistical evidence that \vca\ is linearly
related to stretch.

\citetalias{Maguire12} found a 3.4-$\sigma$ linear relation between
their \vcaz\ measurements and stretch.  Above, our measured
significance is much lower.  This is partly because in the above
calculation, we did not include SNe~2011by and 2011fe since we do not
have the \citetalias{Maguire12} spectra.  Including the reported
values for these SNe, there is only a minor change in the
significance, changing the percentage of realizations with positive
slopes to be 98.7 per cent, which is a 2.5-$\sigma$ result.  The other
difference is that above we examine \vca\ instead of \vcaz.
Performing the same analysis as \citetalias{Maguire12} (using \vcaz\
and including SNe~2011by and 2011fe), we find that 99.4 per cent of
the realizations have positive slopes, which is a 2.8-$\sigma$ result.
The difference in significance is likely in the subtleties of fitting
a line.  This practice is not trivial \citep*{Hogg10}, but the
\citet{Kelly07} method is generally a better choice than most options.

We also performed a Kolmogorov--Smirnov (KS) test, splitting the
samples by a stretch of 1.01.  This is not an ideal test since there
are several SNe with stretches consistent with 1.01 (and therefore
could be in either group) and uncertainty in the velocity can also
change the overall distribution, but it can provide an indication of a
difference.  The KS test resulted in $p$ values of 0.0014, 0.0036, and
0.080 for the \citetalias{Maguire12}, the \ion{Si}{II} $\lambda 3858$,
and Ca H\&K velocity measurements, respectively.  These tests indicate
that the low and high-stretch subsamples have different parent
populations for both the \ion{Si}{II} $\lambda 3858$ and
\citetalias{Maguire12} velocities.  However, there is no statistical
evidence that the low/high-stretch subsamples have different parent
\vca\ distributions.

Although there is only marginal evidence that there is a linear
relationship between the \citetalias{Maguire12} measurements and
stretch, we find a similar significance of a relationship between
\ion{Si}{II} $\lambda 3858$ velocity and stretch.  Since the \vca\
shows no evidence for a correlation with stretch and the
\citetalias{Maguire12} measurements are correlated with the
\ion{Si}{II} $\lambda 3858$ velocity (and at most weakly correlated
with the Ca H\&K velocity), the physical relationship underlying the
result identified by \citetalias{Maguire12} is likely the correlation
between \ion{Si}{II} $\lambda 3858$ velocity and stretch.  From the
{\small SYNOW} models, this relation is understood as a temperature
effect (and not a real difference in photospheric velocity).

\begin{figure}
\begin{center}
\includegraphics[angle=90,width=3.2in]{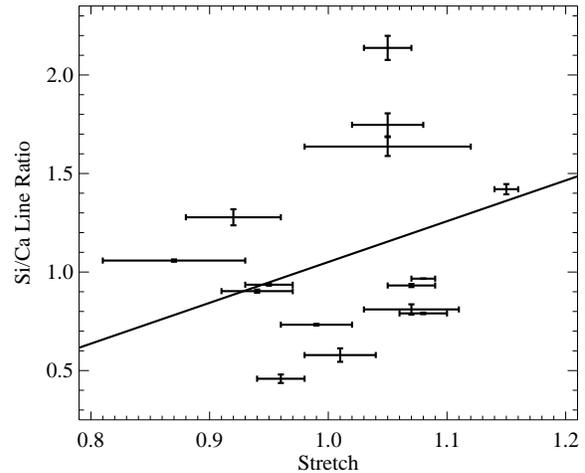}
\caption{Si/Ca ratio as a function of stretch for the
  \citetalias{Maguire12} sample.  The solid line represents the
  best-fitting linear relationship for the data, corresponding to a
  1.4-$\sigma$ result.}\label{f:s_rat}
\end{center}
\end{figure}

Next, we examine the Si/Ca ratio.  We show the Si/Ca ratio as a
function of stretch for the \citetalias{Maguire12} sample in
Figure~\ref{f:s_rat}.  Performing a Bayesian Monte--Carlo linear
regression on the data, we find that 82.4 per cent of the realizations
have positive slopes, corresponding to a 1.4-$\sigma$ result.
Although there is no evidence for a linear relationship between
stretch and the Si/Ca ratio in the \citetalias{Maguire12} data, there
is a slight correlation with higher stretch SNe having larger Si/Ca
ratios.  This is the same general trend expected from the {\small
  SYNOW} models, and future investigations should determine if such a
trend exists.

\begin{figure}
\begin{center}
\includegraphics[angle=90,width=3.2in]{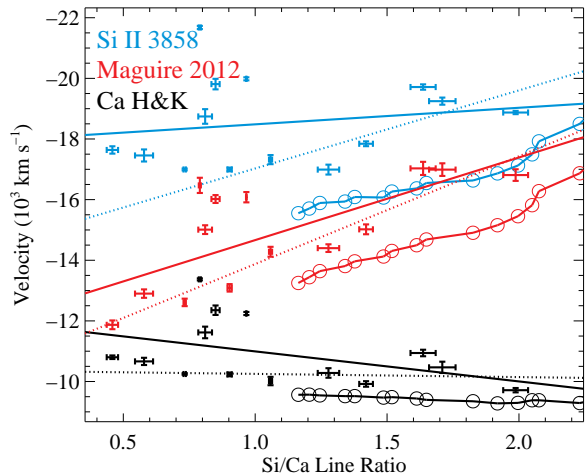}
\caption{Velocities measured with a single or double-Gaussian fit to
  the Ca H\&K feature as a function of the Si/Ca ratio for the
  \citetalias{Maguire12} sample.  The single-Gaussian measurements are
  taken from \citetalias{Maguire12} and represented by the red data.
  The double-Gaussian measurements are shown as black and blue points
  for the Ca H\&K and \ion{Si}{II} $\lambda 3858$ (assuming a rest
  wavelength of 3945~\AA) components, respectively.  The solid lines
  represent the best-fitting linear relationships for the data,
  corresponding to 2.7, 1.0, and 1.7-$\sigma$ results for the
  \citetalias{Maguire12}, \ion{Si}{II} $\lambda 3858$, and Ca H\&K
  velocities, respectively.  The dotted lines represent the
  best-fitting linear relationships for the data with a Si/Ca ratio of
  $>$1.  The red, black, and blue circles represent the
  single-Gaussian and double-Gaussian fit velocities for the {\small
    SYNOW} model spectra.}\label{f:rat_vel}
\end{center}
\end{figure}

Finally, we compare the Si/Ca ratio to our measured velocities
(Figure~\ref{f:rat_vel}).  There are no obvious trends (1.0 and
1.7-$\sigma$) between the Si/Ca ratio and the \ion{Si}{II} $\lambda
3858$ or Ca H\&K velocities.  However, there is a moderate trend
between the Si/Ca ratio and the \citetalias{Maguire12} measurements
(2.7-$\sigma$), where the \citetalias{Maguire12} velocities increase
with increasing Si/Ca ratio.  One should expect that the
single-Gaussian method (as employed by \citetalias{Maguire12}) should
be intermediate to the \ion{Si}{II} $\lambda 3858$ and Ca H\&K
velocities.  The velocity should be closer to the Ca H\&K velocity
when the \ion{Si}{II} $\lambda 3858$ feature is weak (small Si/Ca
ratio) and closer to the \ion{Si}{II} $\lambda 3858$ velocity when the
feature is strong (large Si/Ca ratio).

The behavior seen in the data is reproduced to some extent by the
{\small SYNOW} models.  Figure~\ref{f:rat_vel} also shows the {\small
  SYNOW} model Si/Ca ratio compared to the \ion{Si}{II} $\lambda
3858$, Ca H\&K, and single-Gaussian velocities.  The Ca H\&K velocity
is relatively flat for the {\small SYNOW} models, while the
\ion{Si}{II} $\lambda 3858$ and single-Gaussian velocities increase
with increasing Si/Ca ratio.  The Ca H\&K and single-Gaussian trends
are similar in the data and models, but the data have higher
velocities than the data (by about 700 and 2200~\kms, respectively).
There is no obvious trend in the \ion{Si}{II} $\lambda 3858$ data,
contrary to what is seen in the models.

The models are restricted to Si/Ca ratios of $>$1.  If we fit the data
with the same restriction, the fits are more similar to the slopes
seen in the models.  Specifically, the trend with Ca H\&K is flatter
and the trend with \ion{Si}{II} $\lambda 3858$ is stronger, although
neither trend is statistically significant.  Perhaps a simple linear
relation is not sufficient to describe the trend between the Si/Ca
ratio and \ion{Si}{II} $\lambda 3858$ velocity.


\section{The CfA Sample}\label{s:cfa}

Although the \citetalias{Maguire12} sample was used to identify some
trends which were investigated above, it is limited in size.  To test
additional trends, we use the CfA spectral sample \citep{Blondin12}.
Over the last two decades, the CfA SN Program has observed hundreds of
SNe~Ia, mostly with the FAST spectrograph \citep{Fabricant98} mounted
on the 1.5~m telescope at the F.~L.\ Whipple Observatory.  The data
have been reduced in a consistent manner \citep{Matheson08,
  Blondin12}, producing well-calibrated spectra.  These spectra often
cover Ca H\&K and always cover \ion{Si}{II} $\lambda 6355$, which the
\citetalias{Maguire12} spectra do not cover.

For SNe~Ia in the sample with a measured time of maximum brightness
from light curves, $v_{\rm Si~II}$ and $v_{\rm Ca~H\&K}$ have been
measured \citep{Blondin12}.  Briefly, this is achieved by first
generating a smoothed spectrum using an inverse-variance Gaussian
filter \citep{Blondin06}, and the wavelength of maximum absorption in
the smoothed spectrum is used to determine the velocity (see
\citealt{Blondin12} for details).  The measurements for each spectrum
have been reported by \citet{Foley11:vgrad}, and measurements in all
cases were obtained by \citet{Blondin12}

The CfA sample contains 1630 $v_{\rm Si~II}$ and 1192 $v_{\rm
Ca~H\&K}$ measurements for 255 and 192 SNe~Ia, respectively.

The velocity of each absorption minimum in the \ion{Ca}{II} H\&K
feature of the smoothed spectra is automatically recorded
\citep{Blondin12}.  We only examine spectra with one or two minima.
If two minima are found, the higher/lower velocities are classified as
``blue''/``red.''  If only one minimum is found, it is categorized as
the red or lower-velocity component.

\citet{Foley11:vgrad} noted that comparing all $v_{\rm Ca~H\&K}$
measurements to their corresponding $v_{\rm Si~II}$ measurements,
there were two distinct ``clouds'' corresponding to a lower and higher
velocity relative to $v_{\rm Si~II}$.  The higher-velocity cloud
typically corresponds to the blue velocity component, although there
are some red measurements in that cloud.  The red measurements in the
blue cloud typically have indications of a lower-velocity component,
such as a red shoulder in the line profile, and it was assumed that
they likely corresponded to measurements which were physically similar
to the blue measurements and were simply misclassified.

\begin{figure*}
\begin{center}
\includegraphics[angle=90,width=6.8in]{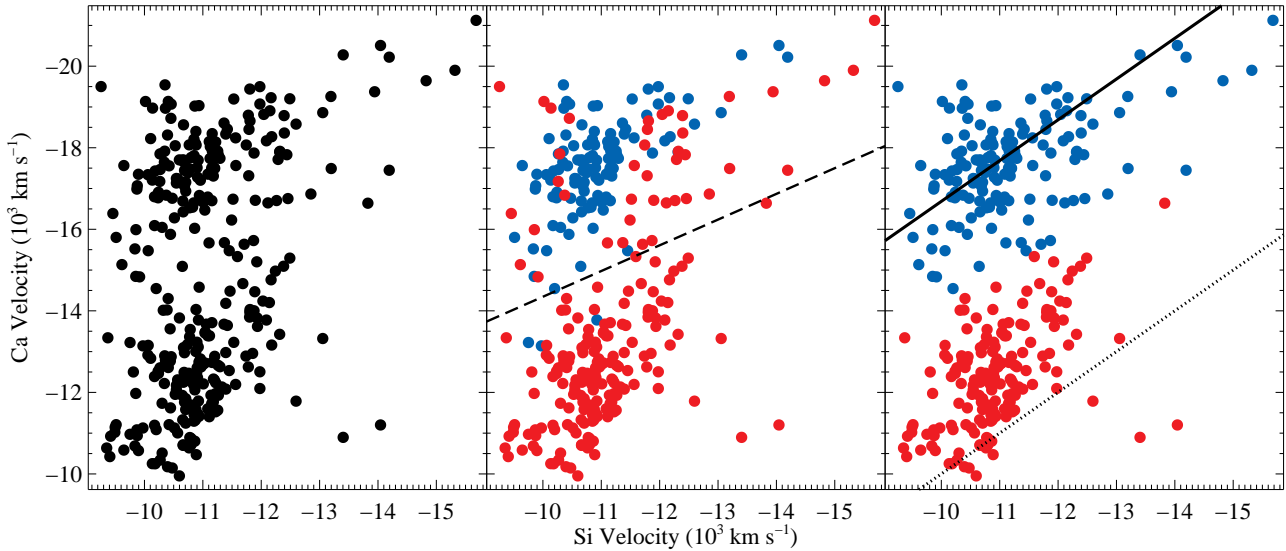}
\caption{Velocities from the Ca H\&K profile (both blue and red
  components), assuming a rest wavelength of 3945~\AA, as a function
  of \ion{Si}{II} $\lambda 6355$ velocity for the CfA sample with a
  phase range of $-1 \le t \le 4.5$~d.  The middle panel plots the
  data as blue and red for the blue and red components, respectively.
  The dashed line represents the line of separation for the two clouds
  from the fitting method of \citet{Williams10}.  The right panel
  removes all measurements from blue components in the red cloud (as
  determined by the dividing line) and assigns all ``red'' velocities
  in the blue cloud as ``blue'' points under the assumption that they
  were originally misclassified.  The dotted line represents spectra
  that would have the same Ca H\&K and \ion{Si}{II} $\lambda 6355$
  velocities.  The solid line represents spectra that would have the
  same Ca H\&K and \ion{Si}{II} $\lambda 6355$ velocities if Ca H\&K
  were actually from \ion{Si}{II} $\lambda 3858$ absorption at the
  same velocity as \ion{Si}{II} $\lambda 6355$.}\label{f:cfa_vel}
\end{center}
\end{figure*}

In Figure~\ref{f:cfa_vel}, we show the subset of CfA measurements of
spectra with $-1 \le t \le 4.5$~d (chosen to match the
\citetalias{Maguire12} sample).  This subset also shows the distinct
blue/red clouds.  We used the method of \citet*{Williams10} to fit a
single slope, but separate offsets to the two clouds.  As a result of
that fitting, there is a natural dividing line between the two clouds,
and we used this line to produce cleaner subsamples.  We removed every
blue measurement in the red cloud, since they may be errant
measurements.  We also reassigned every red measurement in the blue
cloud as a ``blue'' measurement because of the reasons listed above.
This full process is shown graphically in the three panels of
Figure~\ref{f:cfa_vel}.

\begin{figure}
\begin{center}
\includegraphics[angle=90,width=3.2in]{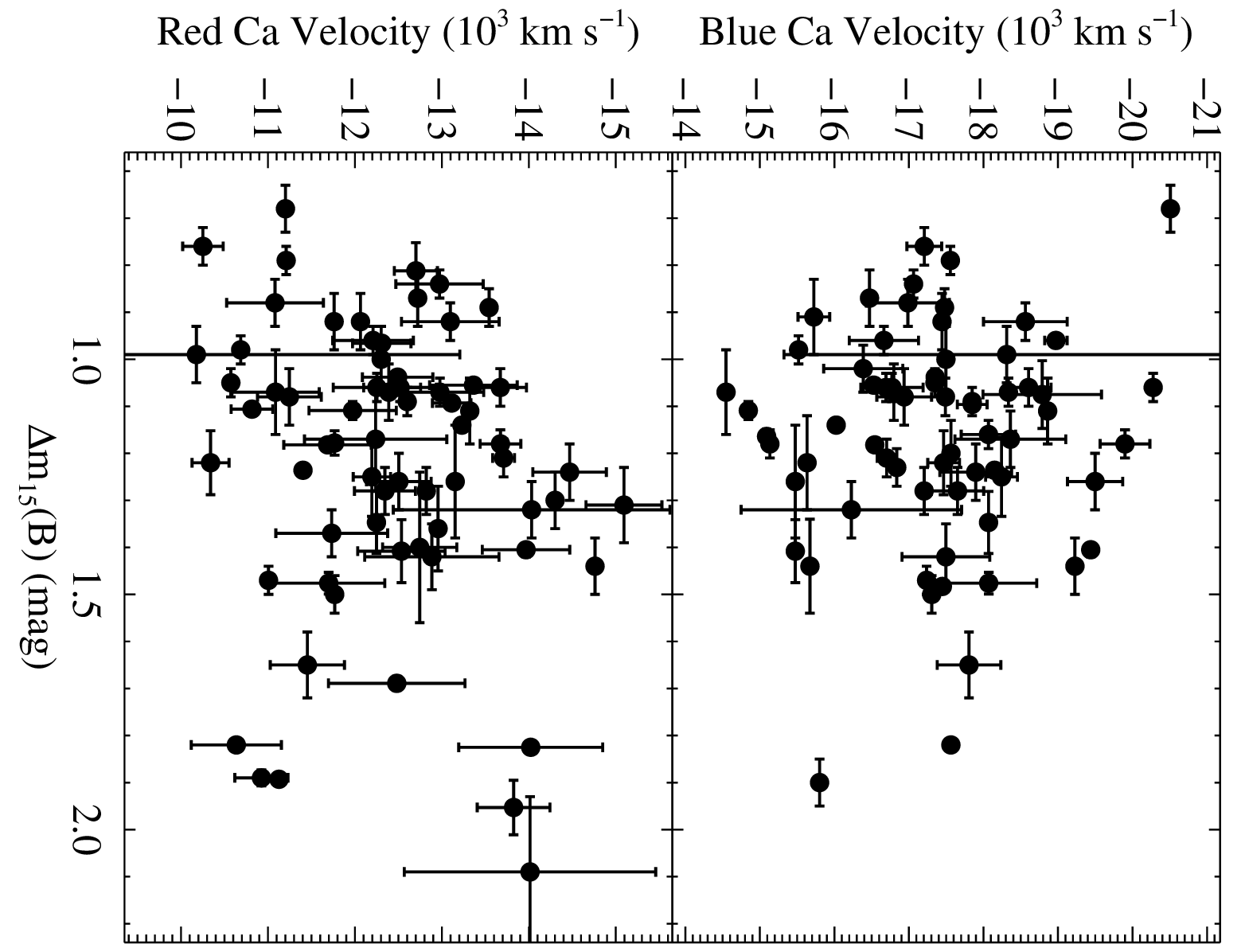}
\caption{Velocity of the blue (top) and red (bottom) components of the
  Ca H\&K profile (assuming a rest wavelength of 3945~\AA) vs.\
  $\Delta m_{15} (B)$ for the CfA sample and a phase range of $-1 \le
  t \le 4.5$~d.  The choice of blue and red velocities were made as
  described in the text and visually represented in
  Figure~\ref{f:cfa_vel}.  Each point represents a different SN (66
  and 67, respectively).  There is no significant correlation for
  either (linear relations are 0.4$\sigma$ and 1.3$\sigma$
  significant, respectively).  Since the red component, corresponding
  to photospheric Ca H\&K, has a slight (but insignificant) trend of
  higher velocity with faster-declining light curves, the data are
  significantly inconsistent with the claim of \citetalias{Maguire12}
  that Ca H\&K velocity decreases with faster-declining light
  curves.}\label{f:vel_dm15}
\end{center}
\end{figure}

With these clean subsets, we have a reasonable estimate of the
velocities for the blue and red components of the Ca H\&K feature.
Since some SNe in the CfA sample have multiple spectra in the chosen
phase range, we created samples for each velocity group where there is
one measurement per SN.  For each SN, we chose the measurement closest
to a phase of $t = 2.7$~d, the median of the \citetalias{Maguire12}
sample.  This resulted in samples of 66 and 67 SNe~Ia with blue and
red measurements (approximately one-third of the full CfA sample and
about 5 times as large as the \citetalias{Maguire12} sample),
respectively.  We present those measurements as a function of
light-curve shape (specifically, $\Delta m_{15} (B)$) in
Figure~\ref{f:vel_dm15}.  There is no significant linear relation
between light-curve shape and the individual velocity components.  In
fact, the stronger relationship of the red velocity component, which
is the best representation of the Ca H\&K photospheric velocity, is a
1.3$\sigma$ result in the \emph{opposite} direction than the
\citetalias{Maguire12} relation (i.e., higher velocity for
slower-declining SNe~Ia).  Because of this opposing trend, the CfA
data are significantly inconsistent with the \citetalias{Maguire12}
relation.  However, this result is consistent with that of
\citet{Foley11:vgrad} and \citet{Foley12:vel} for both Ca H\&K and
\ion{Si}{II} $\lambda 6355$.

In addition to the arguments detailed above, there is additional
evidence in the CfA data that suggest that the blue component of the
Ca H\&K feature is the result of \ion{Si}{II} $\lambda 3858$.  If the
blue component is predominantly from HV Ca H\&K, then one would not
expect a particularly high correlation between its velocity and that
of \ion{Si}{II} $\lambda 6355$.  That is, the velocity of a HV calcium
component could be independent of the photospheric silicon velocity.
However, there is a reasonable correlation (correlation coefficient of
0.54) between the two.  On the other hand, the velocities of the red
and blue components are barely correlated (correlation coefficient of
0.27).  Therefore, the velocity of the blue component has a larger
association with the photospheric velocity of silicon than the
photospheric velocity of calcium.

Perhaps the best evidence that the blue component is from \ion{Si}{II}
$\lambda 3858$ absorption is presented in Figure~\ref{f:cfa_vel}.  In
the right panel, we plot the relation between \vsi\ and \vca\ in the
scenario where the \vca\ measurement is from a misidentified
\ion{Si}{II} $\lambda 3858$ feature at the same velocity as
\ion{Si}{II} $\lambda 6355$.  The line goes directly through the blue
cloud, indicating that the blue component of the Ca H\&K feature has a
velocity consistent with that of \ion{Si}{II} $\lambda 6355$ if it is
formed by \ion{Si}{II} $\lambda 3858$ absorption.  In other words, the
blue component is at the wavelength one expects by blueshifting
3858~\AA\ by \vsi.  Addtionally, \citet{Blondin12} presented several
examples of SNe where the blue component was consistent with
\ion{Si}{II} $\lambda 3858$ at the \ion{Si}{II} $\lambda 6355$
velocity, while the red component was consistent with the velocity of
the Ca NIR triplet.

From the large CfA sample, we showed additional evidence that \vca\
does not correlate with light-curve shape.  The velocity of the blue
component is correlated with the photospheric silicon velocity (as
measured by \ion{Si}{II} $\lambda 6355$) and relatively uncorrelated
with the photospheric calcium velocity.  In addition to being
correlated with photospheric silicon velocity, the CfA data show that
the velocity of the blue component matches the expected photospheric
velocity of \ion{Si}{II} $\lambda 3858$.


\section{Additional Modeling}\label{s:add}

From the above analysis of the {\small SYNOW} models, the comparison
of the \citetalias{Maguire12} sample to the {\small SYNOW} models, and
an examination of the CfA sample, there is significant evidence that
the blue component of the Ca H\&K feature is predominantly from
\ion{Si}{II} $\lambda 3858$.  However, additional confidence in this
claim can be obtained by modeling specific SNe.

In this section, we examine the two possible scenarios for the blue
component of the Ca H\&K feature (either HV calcium or \ion{Si}{II}
$\lambda 3858$) for two test cases: SN~2011fe and SN~2010ae.

\subsection{SN~2011fe}\label{ss:11fe}

SN~2011fe, which occurred in M~101 and was the brightest SN~Ia in 40
years, has been incredibly well observed and extensively studied
\citep[e.g.,][]{Nugent11, Brown12, Chomiuk12, Horesh12, Margutti12,
  Matheson12, Parrent12, Shappee13}.  Here we examine a single
maximum-light spectrum of SN~2011fe, obtained by \emph{HST} using the
STIS spectrograph (Program GO--12298; PI Ellis).  The spectra were
obtained on 2011 September 10 between 09:51 and 11:14 UT,
corresponding to $t = 0.0$~d relative to $B$-band maximum brightness
\citepalias{Maguire12}.  The observations were obtained with three
different gratings, all with the $52\arcsec \times 0.\arcsec2$ slit.
Two exposures were obtained for each of the CCD/G230LB, CCD/G430L, and
CCD/G750L setups with individual exposure times of 530, 80, and 80~s,
respectively.  The three setups yield a combined wavelength range of
1665 -- 10,245~\AA.  The data were reduced using the standard
\textit{HST} Space Telescope Science Data Analysis System (STSDAS)
routines to bias subtract, flat-field, extract, wavelength-calibrate,
and flux-calibrate each SN spectrum.

\begin{figure*}
\begin{center}
\includegraphics[angle=90,width=6.8in]{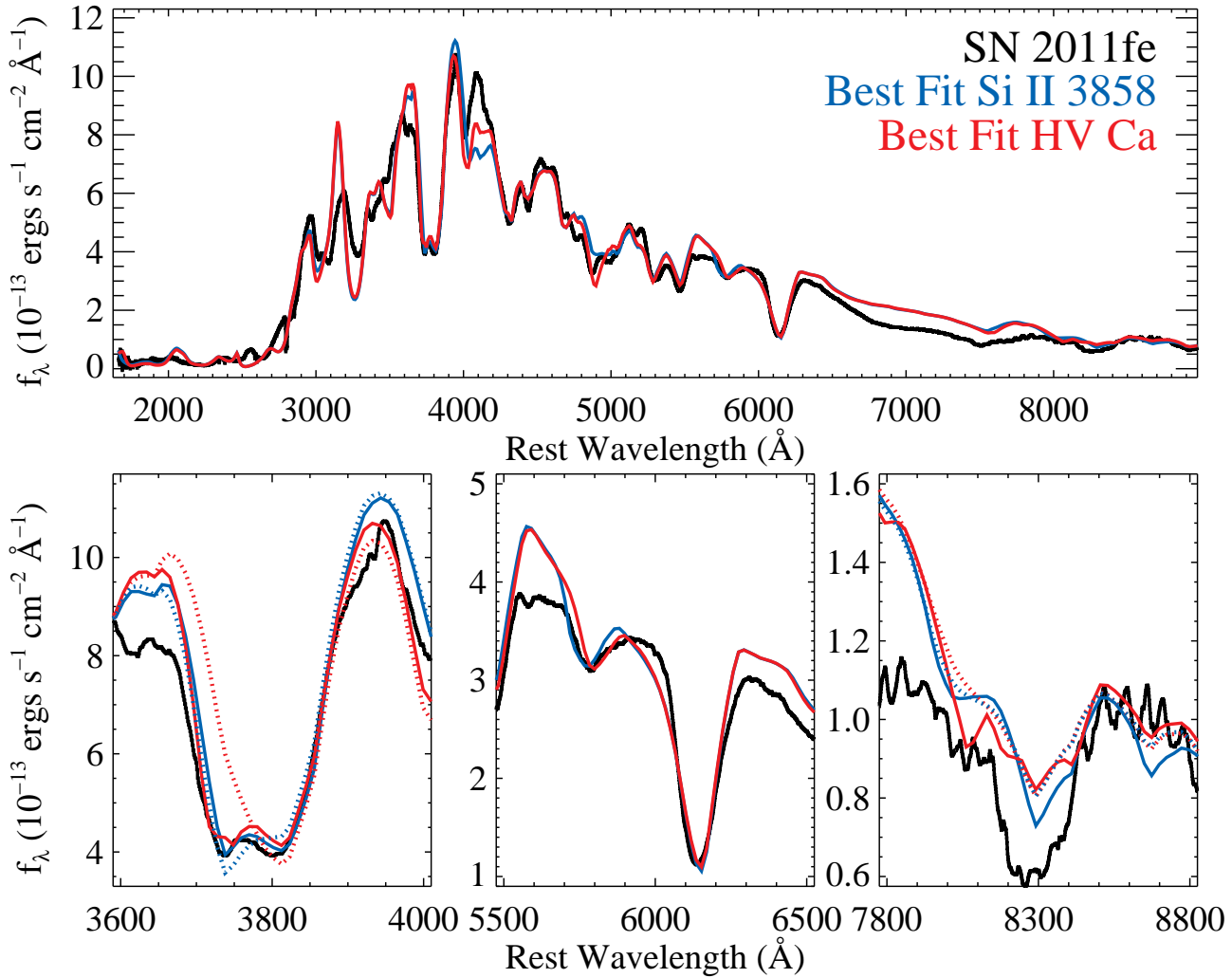}
\caption{\emph{HST} spectrum of SN~2011fe at $t = 0.0$~d (black
  curve).  The blue and red curves are ``best-fitting'' {\small SYNOW}
  model spectra where there was an attempt to simultaneously match the
  Ca H\&K profile and \ion{Si}{II} $\lambda 5972$ and $\lambda 6355$
  lines using \ion{Si}{II}/HV \ion{Ca}{II} to match the blue component
  of the Ca H\&K feature, respectively, and differences in the
  strength of Na~D to match the \ion{Si}{II} $\lambda 5972$ feature.
  The dotted red line represents the model represented by the solid
  red line except without a HV \ion{Ca}{II} component.  The dotted
  blue line represents the model represented by the solid blue line,
  except with the \ion{Ca}{II} opacity and density structure matched
  to that of the model represented by the solid red
  line.}\label{f:11fe}
\end{center}
\end{figure*}

We present the spectrum in Figure~\ref{f:11fe}.  We note that
\citetalias{Maguire12} presented a spectrum for SN~2011fe from a
different phase and which only covered \about 2900 -- 5700~\AA.  This
is the first publication of these data.  This is also only the second
published maximum-light SN~Ia spectrum to probe below \about 2500~\AA\
(the first being of SN~2011iv; \citealt{Foley12:11iv}).

The SN~2011fe spectrum is of extreme high quality, including in the
UV.  Because of its quality and wavelength coverage, we can produce a
reasonable {\small SYNOW} model.  We have made two attempts at fitting
the SN~2011fe spectrum using {\small SYNOW} models.  The first assumes
that the blue component of the Ca H\&K feature is from \ion{Si}{II}
$\lambda 3858$; the second assumes that it is caused by HV Ca H\&K.
We present our models in Figure~\ref{f:11fe} and model parameters in
Table~\ref{t:models}.

When generating these models, we first attempted to fit the full
spectrum with a limited number of species.  These models are not
optimized to fit the entire spectrum; because of potential effects
other species could have on the spectral features of interest, we
wanted a first-order model of the full spectrum.  We then either added
HV \ion{Ca}{II} or adjusted the \ion{Si}{II} temperature to match the
blue component of the Ca H\&K feature.  We allow the opacity and
density structures for \ion{Ca}{II}, \ion{Si}{II}, HV \ion{Ca}{II},
and \ion{Na}{I} to vary, but all other species remain the same.

For the HV calcium model, we adjusted the \ion{Si}{II} temperature to
an extreme value that still fits the \ion{Si}{II} $\lambda 5972$
feature.  In this model, we do not include any \ion{Na}{I}, and
therefore Na~D does not contribute at all to this feature.  As a
result, the \ion{Si}{II} $\lambda 3858$ is about as weak as possible,
and the HV Ca H\&K is essentially as strong as possible.

For the \ion{Si}{II} $\lambda 3858$ model, we adjust the \ion{Si}{II}
temperature to an extreme value to match the blue feature in the Ca
H\&K feature.  We then add \ion{Na}{I} to match the strength of the
feature near 5800~\AA.

These models differ in some ways from those presented by
\citet{Parrent12} for their optical-only maximum-light SN~2011fe
spectrum.  Most differences are related to matching the UV region,
which requires adding \ion{Co}{II} and \ion{Cr}{II}.  Interestingly,
adding these features reduces the need to include \ion{Fe}{II} in the
{\small SYNOW} model (although we cannot definitively say that it is
not in the spectrum).  Additionally, we are able to better model the
Ca H\&K feature than \citet{Parrent12} because of the additional data
blueward of the feature.

Examining the {\small SYNOW} models in detail, particularly near the
Ca H\&K feature, the redder \ion{Si}{II} features, and the Ca NIR
triplet (see lower panels of Figure~\ref{f:11fe}), we see that the
models are very similar.  In other words, {\small SYNOW} modeling of
SN~2011fe cannot distinguish between our two scenarios; it simply has
too many parameters for the data.

We did not adjust the models to fit the Ca NIR triplet, with the hope
that we might see signatures of HV Ca.  There is a feature in the
{\small SYNOW} model that is coincident with a shoulder in the
SN~2011fe spectrum.  However, we see a similar feature in the
\ion{Si}{II} $\lambda 3858$ model that is simply the result of a
slightly different density profile for \ion{Ca}{II}.  A full spectral
sequence and/or NIR spectra, which would supply additional
\ion{Si}{II} features, may provide a clear way to distinguish the
models.

\subsection{SN~2010ae}

With an inconclusive result from modeling SN~2011fe, we now turn to
modeling SN~2010ae.  SN~2010ae is a SN~Iax \citep{Foley12:iax} similar
to SN~2008ha \citep{Foley09:08ha, Foley10:08ha, Valenti09}. Its
spectrum is similar to that of a SN~Ia, but with an extremely low
ejecta velocity.  This indicates that the ejecta composition, density
structure, temperature, and other aspects of the explosion important
for producing a particular SED are similar for SN~2010ae and SNe~Ia.
However, because of the low ejecta velocity, line blending is minimal.

\begin{figure*}
\begin{center}
\includegraphics[angle=90,width=6.8in]{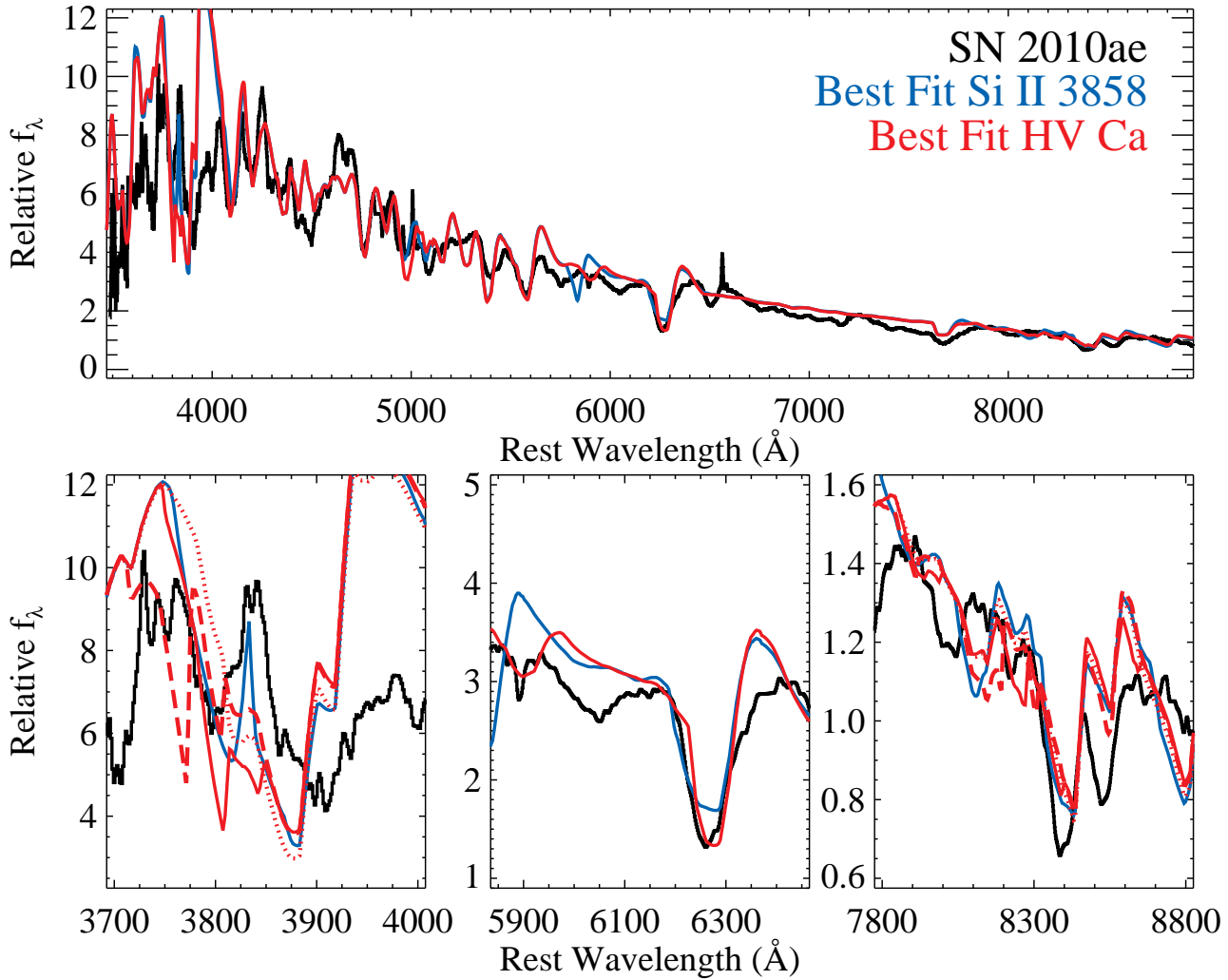}
\caption{Same as Figure~\ref{f:11fe}, except for SN~2010ae and its
  ``best-fitting'' {\small SYNOW} models.  The SN~2010ae spectrum has
  been smoothed slightly and dereddened as described in the text.  The
  solid and dashed red lines are for HV calcium where Ca~K and Ca~H,
  respectively, are matched to the absorption near
  3800~\AA.}\label{f:10ae}
\end{center}
\end{figure*}

We present a near maximum-light spectrum of SN~2010ae originally
presented by \citet{Foley12:iax} and presumed to be obtained near
maximum light in Figure~\ref{f:10ae}.  This spectrum only covers
optical wavelengths.  We dereddened the spectrum by $E(B-V) = 0.6$~mag
to roughly match the continuum of SN~2011fe and smoothed the spectrum
with a inverse-variance weighted Gaussian filter and velocity scale of
150~\kms.

Perhaps the most important aspect of the SN~2010ae spectrum is that
the Ca H\&K feature is separated into two distinct features.

We then attempted to produce {\small SYNOW} model spectra in a way
similar to what was performed for SN~2011fe.  As a starting point, we
used the SN~2011fe models.  We decreased $v_{\rm phot}$ from
9000~\kms\ to 3000~\kms.  We also reduced minimum and maximum
velocities for each species.  The details of the models are presented
in Table~\ref{t:models}.

We did not change the majority of parameters for the model.  As a
result, the fits are not ideal.  In particular, the lack of
\ion{C}{II} results in missing obvious features.  Additional
adjustments would certainly improve the overall fit, but this is not
necessary for our purpose.  However, keeping the model similar to that
of a SN~Ia (with mostly just adjustments to the velocity) reinforces
the spectral (and compositional) similarities between SNe~Iax and
SNe~Ia.

Additionally, we changed the opacity of \ion{Si}{II} and \ion{Ca}{II},
and we changed the density structure of \ion{Ca}{II}.  We adjusted the
opacity of \ion{Si}{II} to roughly match the \ion{Si}{II} $\lambda 6355$
feature.  The \ion{Ca}{II} opacity was changed to roughly match the NIR
triplet.  The velocity of the HV calcium and the density structure of
both the HV and photospheric calcium were adjusted to match the Ca
H\&K feature.

Ca H\&K are offset by 34.8~\AA, which corresponds to 2640~\kms.  The
velocity difference between the two components will be present even if
Ca H\&K are blueshifted.  For most SNe, the ejecta velocities are high
enough where the two components blend together completely.  But for
SN~2010ae, which has an ejecta velocity similar to this separation,
any Ca H\&K feature will be roughly twice the width of a feature from
a single line.  For SN~2010ae, the blue component of the Ca H\&K
feature has a FWHM of 2960~\kms.  Therefore, the Ca H\&K components
can barely fit within the width of the feature (with a velocity of
\about 11200~\kms, about 4 times that of the photospheric velocity),
but then the line can only be minimally broadened.  That is
unphysical, but if it were the case, then one would expect two
components within the blue component, which is not seen.

The only other choice is to choose a velocity which results in either
Ca~H or Ca~K to have a minimum near 3800~\AA.  Doing this for Ca~H
results in a velocity of \about 13000~\kms\ and a significant
absorption feature at \about 3760~\AA, where no such feature exists.
When assigning a velocity of \about 10000~\kms\ for HV calcium (such
that Ca~K is at \about 3800~\AA), there is no gap between the blue and
red components.  Neither option reproduces the observed profile for
SN~2010ae.

Alternatively, the \ion{Si}{II} $\lambda 3858$ model roughly matches
the spectrum of SN~2010ae.  In particular, it reproduces the (now
unblended) Ca H\&K feature.  The HV calcium model, on the other hand,
does not reproduce a key aspect of the Ca H\&K feature -- its
unblended nature.  It is reasonably certain that \ion{Si}{II} $\lambda
3858$ causes the absorption of the blue component of the normally
blended Ca H\&K feature for SN~2010ae.

Furthermore, removing the HV Ca from the HV Ca model does not have two
distinct features.  It appears necessary to have a reasonably strong
\ion{Si}{II} $\lambda 3858$ feature to produce the emission between
the two components.

Since \citet{Foley12:iax} showed that SNe~Iax have very similar
spectra to SNe~Ia, except with different velocities, and since the
SN~2011fe {\small SYNOW} model roughly matches the SED of SN~2010ae
(with only differences in the velocity), one can extrapolate this
result to SNe~Ia.


\section{Discussion \& Conclusions}\label{s:conc}

We have shown through a re-examination of the \citetalias{Maguire12}
sample, a re-examination of the CfA sample, basic {\small SYNOW}
modeling, and more thorough {\small SYNOW} modeling of SNe~2010ae and
2011fe that the blue component of the Ca H\&K spectral feature in
near-maximum light SN~Ia spectra is typically from \ion{Si}{II}
$\lambda 3858$ absorption.  This was also the interpretation of
\citet{Wang03}, which has spectropolarimetric observations of Ca H\&K,
\ion{Si}{II} $\lambda 6355$, and the Ca NIR triplet, providing
additional weight to this conclusion.  Some previous claims that the
component is the result of HV Ca H\&K absorption may require
re-examination.  The Ca NIR triplet has shown HV features for some
SNe, although it is also possible to reproduce some of these features
with a different (but still smooth) density profile for calcium (see
Section~\ref{ss:11fe}).  Therefore, it is still unclear if HV calcium
contributes to the Ca H\&K component, how frequently it does, and if
that contribution is typically blended with \ion{Si}{II} $\lambda
3858$.  The realization that the blue absorption in the Ca H\&K
profile is from \ion{Si}{II} $\lambda 3858$ for most SNe~Ia has
far-reaching implications for our understanding of SN~Ia progenitor
systems and explosion models, which have interpreted the prevalence of
HV calcium as an indication of specific explosion mechanisms and
potentially a tracer of the environment of the progenitor system.

Because the Ca H\&K profile is a combination of Ca H\&K and
\ion{Si}{II} $\lambda 3858$, \vca\ should not be measured by fitting
the entire Ca H\&K feature with a single (Gaussian) component.
Regardless of the source of the two components, we also show that if
one \emph{does} fit the profile with a single Gaussian component that
the resulting measurements will be unphysical, inaccurate, and highly
biased.  However, because of the true nature of the blue component, a
single Gaussian fit is particularly biased.

We confirmed the \citetalias{Maguire12} result that SNe in their
sample have different Ca H\&K line profiles based on light-curve
shape.  However, the difference is mostly constrained to the blue
component, with no evidence for a difference in velocity or width for
the red component.

We re-examined the claim that \vcaz\ is correlated with light-curve
shape \citepalias{Maguire12}.  Using the reported
\citetalias{Maguire12} measurements, we do not find a statistically
significant linear relation, but the KS test does indicate different
parent populations for low/high-stretch subsamples.  When using \vca\
measurements from the red component of the Ca H\&K profile for the
\citetalias{Maguire12} spectra, there is no statistically significant
trend between \vca\ and light-curve shape.  An analysis of the CfA
sample also showed that there is no correlation between ejecta
velocity and light-curve shape, confirming the previous results of
\citet{Foley11:vgrad} and \citet{Foley12:vel}.  Instead, the
underlying physical effect driving the relation between the
\citetalias{Maguire12} measurements and light-curve shape is likely
the relation between \ion{Si}{II} $\lambda 3858$ and temperature.

This result implies that the \citetalias{Maguire12} claim that \vcaz\
does not correlate with host-galaxy mass is not supported by data.
Other claims made by \citetalias{Maguire12} related to \vcaz,
including correlations between \vcaz\ and the wavelengths or
velocities of certain features, should also be re-examined.

From modeling, there is some indication that the Si/Ca ratio should be
a strong tracer of temperature and an indicator of light-curve shape,
but this is not verified with data.  There may also be a relatively
low correlation between \vca\ and the pseudo-equivalent width of the
Ca H\&K feature.  This may be why \citet{Foley11:vgrad} did not find a
relation between the pseudo-equivalent width of the Ca H\&K feature
and the intrinsic colour of SNe~Ia.

\citet{Foley11:vgrad} and \citet{Foley12:vel} suggested that \vcaz\
could be useful for measuring the intrinsic colour of SNe~Ia.
However, this current analysis shows that this approach may be limited
by the contamination of \ion{Si}{II} $\lambda 3858$.  At the very
least, SNe with very high ejecta velocities will have a Ca H\&K
profile that is a blend of \ion{Si}{II} $\lambda 3858$ and Ca H\&K
with no distinct components.  At that point, one should be circumspect
of the derived velocity.  The culling technique of
\citet{Foley11:vgrad} should reduce the number of spectra with
velocity measurements contaminated by \ion{Si}{II} $\lambda 3858$, but
relatively low signal-to-noise ratio (S/N) spectra, galaxy
contamination, and other nuisances, may reduce the viability of this
option.

There is a proposal to have a low-resolution ($R \approx 75$)
spectrograph on WFIRST \citep{Green12}.  The main purpose of the
spectrograph for SN science would be spectroscopic classification and
redshift determination.  Similarly, the SED Machine \citep{Ben-Ami12},
is a proposed $R \approx 100$ spectrograph to classify thousands of
low-redshift SNe.  Another use of these spectrographs could be to
measure ejecta velocities.  Assuming perfect knowledge of the SN
redshift, the precision of the ejecta velocity measurement can be
limited by spectroscopic resolution.

\begin{figure}
\begin{center}
\includegraphics[angle=90,width=3.2in]{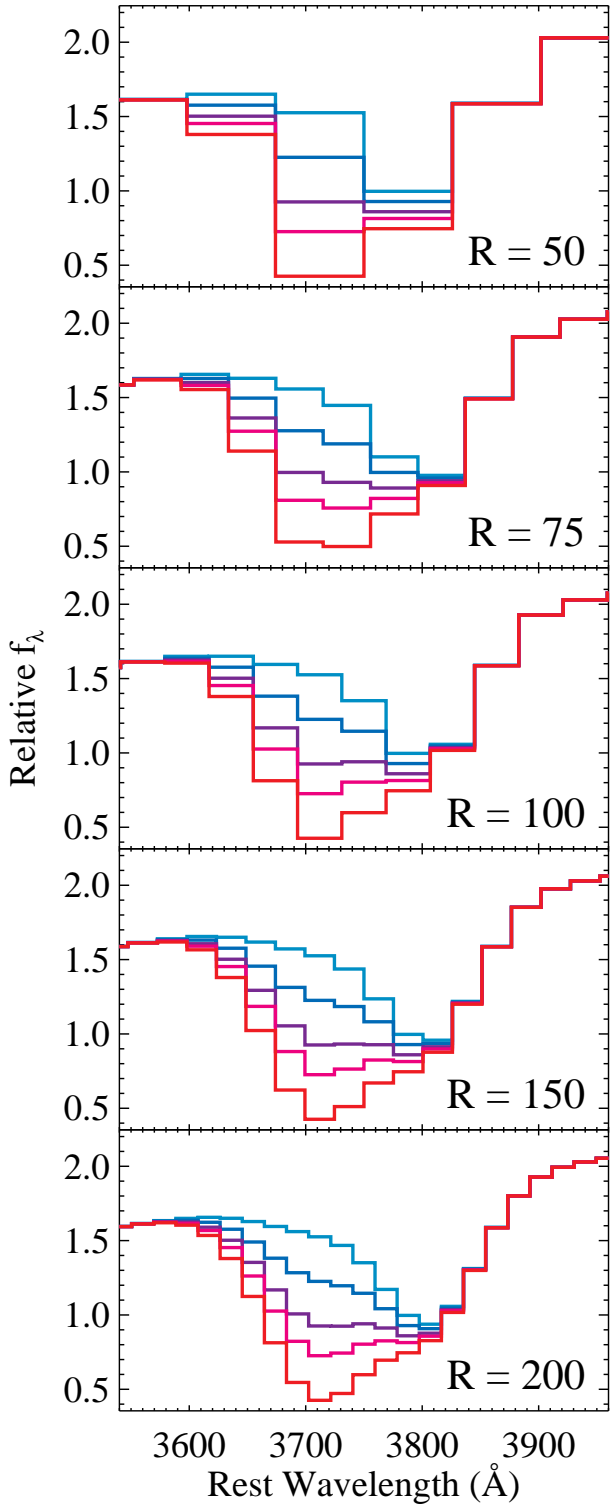}
\caption{Artificial Ca H\&K line profiles.  The profiles are the same
  as shown in Figure~\ref{f:gauss}.  Different resolutions ($R = 50$,
  75, 100, 150, and 200) are shown in each panel from top to
  bottom.}\label{f:test_res}
\end{center}
\end{figure}

To test our ability to determine ejecta velocities with different
resolutions, we show artificial Ca H\&K line profiles that contain two
components in Figure~\ref{f:test_res}.  One cannot distinguish the two
components of the profile at $R = 50$; there are \about 4 resolution
elements in the feature, which is insufficient for a full
six--parameter fit of a double-Gaussian fit.  Additionally, the two
components are separated by \about 6000~\kms, corresponding to $R
\approx c / 6000{\rm ~km~s}^{-1} \approx 50$.  At $R = 75$, one can
start to see the effect of the two components in some spectra (i.e.,
flat bottoms), but the components are still not clearly separate.  A
resolution of 100 may be the minimal amount to clearly see the effects
of multiple components.  But considering additional effects such as
potential [\ion{O}{II}] $\lambda 3727$ emission from the host galaxy
contaminating the line profile, one might want a higher resolution,
such as $R = 200$, where narrow lines should not significantly affect
the overall profile shape.

However, we note that \ion{Si}{II} $\lambda 6355$ does not suffer these
same problems, and $R = 75$ should provide accurate (and reasonably
precise) measurements of the ejecta velocity.  For optical
spectrographs, one can easily measure \vsi\ to $z = 0.3$.  With
red-sensitive CCDs and good sky subtraction one can use optical
spectrographs to measure \vsi\ to $z \approx 0.6$.  With NIR
spectrographs, one can easily measure \vsi\ to $z \approx 2$
(neglecting the faintness of the SNe).

For the SED Machine, which aims to classify low-redshift SNe, it
should also be able to measure \vsi.  The proposed spectrograph on
WFIRST would have a wavelength range of 0.6 -- 2~$\mu$m, which should
cover \vsi\ to $z \approx 3$, well beyond the expected redshift range
of WFIRST.  \citet{Rodney12} presented an {\it HST} observer-frame NIR
spectrum of a $z = 1.55$ SN~Ia, SN Primo.  The spectrum has a low S/N
and is low-resolution ($R \approx 130$).  But using the method of
\citet{Blondin06}, we measure $v_{\rm Si~II} = -11200 \pm 900$~\kms\
at a phase of $6 \pm 3$~d, corresponding to $v_{\rm Si~II}^{0} =
-11700 \pm 1000$~\kms.  This corresponds, using the
\citet{Foley11:vgrad} relations, to $(B_{\rm max} - V_{\rm max})_{0} =
0.00 \pm 0.07$~mag.  The uncertainty in the velocity measurement is
dominated by the low S/N of the spectrum, but the uncertainty in the
intrinsic colour is still dominated by the uncertainty and scatter in
the velocity-colour relation.  None the less, SN Primo appears to be
have a moderate intrinsic colour.  This shows the potential of using
velocity measurements for SN~Ia cosmology even if the complexities of
the Ca H\&K profile prevents accurate measurements.

The additional knowledge of the Ca H\&K profile provided here is a
step toward further understanding of the full SED of SNe~Ia.  SNe~Iax,
which have compositions similar to that of SNe~Ia, can be exceedingly
useful for determining which specific atomic transitions contribute to
SN~Ia spectra.  Because of their low ejecta velocities, SNe~Iax may
provide additional insight into the specific contributions from
various lines for blended SN~Ia features.  Similarly, additional
spectropolarimetric observations of SNe~Ia, and particularly those
that cover both Ca H\&K and the Ca NIR triplet, NIR spectra, and good
spectral sequences starting at early times should produce additional
insight into the formation of a SN~Ia SED.

\section*{Acknowledgments}

  {\it Facilities:} HST(STIS)

  \bigskip We thank D.\ Kasen, R.\ Kirshner, and J.\ Parrent for their
  comments, insights, and help.

  Based on observations made with the NASA/ESA Hubble Space Telescope,
  obtained from the data archive at the Space Telescope Science
  Institute.  STScI is operated by the Association of Universities for
  Research in Astronomy, Inc.\ under NASA contract NAS 5--26555.


\bibliographystyle{mn2e}
\bibliography{../astro_refs}

\begin{thebibliography}{66}
\expandafter\ifx\csname natexlab\endcsname\relax\def\natexlab#1{#1}\fi

\bibitem[{{Altavilla} {et~al}\mbox{.}(2007){Altavilla}, {Stehle},
  {Ruiz-Lapuente}, {Mazzali}, {Pignata}, {Balastegui}, {Benetti}, {Blanc},
  {Canal}, {Elias-Rosa}, {Goobar}, {Harutyunyan}, {Pastorello}, {Patat},
  {Rich}, {Salvo}, {Schmidt}, {Stanishev}, {Taubenberger}, {Turatto}, \&
  {Hillebrandt}}]{Altavilla07}
{Altavilla} G. {et~al.}, 2007, \aap, 475, 585

\bibitem[{{Ben-Ami} {et~al}\mbox{.}(2012){Ben-Ami}, {Konidaris}, {Quimby},
  {Davis}, {Ngeow}, {Ritter}, \& {Rudy}}]{Ben-Ami12}
{Ben-Ami} S., {Konidaris} N., {Quimby} R., {Davis} J.~T., {Ngeow} C.~C.,
  {Ritter} A., {Rudy} A., 2012, in Society of Photo-Optical Instrumentation
  Engineers (SPIE) Conference Series, Vol. 8446, Society of Photo-Optical
  Instrumentation Engineers (SPIE) Conference Series

\bibitem[{{Blondin} {et~al}\mbox{.}(2006){Blondin}, {Dessart}, {Leibundgut},
  {Branch}, {H{\"o}flich}, {Tonry}, {Matheson}, {Foley}, {Chornock},
  {Filippenko}, {Sollerman}, {Spyromilio}, {Kirshner}, {Wood-Vasey},
  {Clocchiatti}, {Aguilera}, {Barris}, {Becker}, {Challis}, {Covarrubias},
  {Davis}, {Garnavich}, {Hicken}, {Jha}, {Krisciunas}, {Li}, {Miceli},
  {Miknaitis}, {Pignata}, {Prieto}, {Rest}, {Riess}, {Salvo}, {Schmidt},
  {Smith}, {Stubbs}, \& {Suntzeff}}]{Blondin06}
{Blondin} S. {et~al.}, 2006, \aj, 131, 1648

\bibitem[{{Blondin} {et~al}\mbox{.}(2012{\natexlab{a}}){Blondin}, {Dessart},
  {Hillier}, \& {Khokhlov}}]{Blondin13}
{Blondin} S., {Dessart} L., {Hillier} D.~J., {Khokhlov} A.~M.,
  2012{\natexlab{a}}, ArXiv e-prints, 1211.5892

\bibitem[{{Blondin} {et~al}\mbox{.}(2012{\natexlab{b}}){Blondin}, {Matheson},
  {Kirshner}, {Mandel}, {Berlind}, {Calkins}, {Challis}, {Garnavich}, {Jha},
  {Modjaz}, {Riess}, \& {Schmidt}}]{Blondin12}
{Blondin} S. {et~al.}, 2012{\natexlab{b}}, \aj, 143, 126

\bibitem[{{Bongard} {et~al}\mbox{.}(2008){Bongard}, {Baron}, {Smadja},
  {Branch}, \& {Hauschildt}}]{Bongard08}
{Bongard} S., {Baron} E., {Smadja} G., {Branch} D., {Hauschildt} P.~H., 2008,
  \apj, 687, 456

\bibitem[{{Branch} {et~al}\mbox{.}(1985){Branch}, {Doggett}, {Nomoto}, \&
  {Thielemann}}]{Branch85}
{Branch} D., {Doggett} J.~B., {Nomoto} K., {Thielemann} F.-K., 1985, \apj, 294,
  619

\bibitem[{{Branch} {et~al}\mbox{.}(2005){Branch}, {Baron}, {Hall}, {Melakayil},
  \& {Parrent}}]{Branch05}
{Branch} D., {Baron} E., {Hall} N., {Melakayil} M., {Parrent} J., 2005, \pasp,
  117, 545

\bibitem[{{Branch} {et~al}\mbox{.}(2007){Branch}, {Troxel}, {Jeffery},
  {Hatano}, {Musco}, {Parrent}, {Baron}, {Dang}, {Casebeer}, {Hall}, \&
  {Ketchum}}]{Branch07}
{Branch} D. {et~al.}, 2007, \pasp, 119, 709

\bibitem[{{Brown} {et~al}\mbox{.}(2012){Brown}, {Dawson}, {de Pasquale},
  {Gronwall}, {Holland}, {Immler}, {Kuin}, {Mazzali}, {Milne}, {Oates}, \&
  {Siegel}}]{Brown12}
{Brown} P.~J. {et~al.}, 2012, \apj, 753, 22

\bibitem[{{Chomiuk} {et~al}\mbox{.}(2012){Chomiuk}, {Soderberg}, {Moe},
  {Chevalier}, {Rupen}, {Badenes}, {Margutti}, {Fransson}, {Fong}, \&
  {Dittmann}}]{Chomiuk12}
{Chomiuk} L. {et~al.}, 2012, \apj, 750, 164

\bibitem[{{Chornock} \& {Filippenko}(2008)}]{Chornock08}
{Chornock} R., {Filippenko} A.~V., 2008, \aj, 136, 2227

\bibitem[{{Conley} {et~al}\mbox{.}(2011){Conley}, {Guy}, {Sullivan},
  {Regnault}, {Astier}, {Balland}, {Basa}, {Carlberg}, {Fouchez}, {Hardin},
  {Hook}, {Howell}, {Pain}, {Palanque-Delabrouille}, {Perrett}, {Pritchet},
  {Rich}, {Ruhlmann-Kleider}, {Balam}, {Baumont}, {Ellis}, {Fabbro},
  {Fakhouri}, {Fourmanoit}, {Gonz{\'a}lez-Gait{\'a}n}, {Graham}, {Hudson},
  {Hsiao}, {Kronborg}, {Lidman}, {Mourao}, {Neill}, {Perlmutter}, {Ripoche},
  {Suzuki}, \& {Walker}}]{Conley11}
{Conley} A. {et~al.}, 2011, \apjs, 192, 1

\bibitem[{{Fabricant} {et~al}\mbox{.}(1998){Fabricant}, {Cheimets}, {Caldwell},
  \& {Geary}}]{Fabricant98}
{Fabricant} D., {Cheimets} P., {Caldwell} N., {Geary} J., 1998, \pasp, 110, 79

\bibitem[{{Fisher} {et~al}\mbox{.}(1997){Fisher}, {Branch}, {Nugent}, \&
  {Baron}}]{Fisher97}
{Fisher} A., {Branch} D., {Nugent} P., {Baron} E., 1997, \apjl, 481, L89+

\bibitem[{{Foley} {et~al}\mbox{.}(2009){Foley}, {Chornock}, {Filippenko},
  {Ganeshalingam}, {Kirshner}, {Li}, {Cenko}, {Challis}, {Friedman}, {Modjaz},
  {Silverman}, \& {Wood-Vasey}}]{Foley09:08ha}
{Foley} R.~J. {et~al.}, 2009, \aj, 138, 376

\bibitem[{{Foley} {et~al}\mbox{.}(2010){Foley}, {Brown}, {Rest}, {Challis},
  {Kirshner}, \& {Wood-Vasey}}]{Foley10:08ha}
{Foley} R.~J., {Brown} P.~J., {Rest} A., {Challis} P.~J., {Kirshner} R.~P.,
  {Wood-Vasey} W.~M., 2010, \apjl, 708, L61

\bibitem[{{Foley} \& {Kasen}(2011)}]{Foley11:vel}
{Foley} R.~J., {Kasen} D., 2011, \apj, 729, 55

\bibitem[{{Foley} {et~al}\mbox{.}(2011){Foley}, {Sanders}, \&
  {Kirshner}}]{Foley11:vgrad}
{Foley} R.~J., {Sanders} N.~E., {Kirshner} R.~P., 2011, \apj, 742, 89

\bibitem[{{Foley}(2012)}]{Foley12:vel}
{Foley} R.~J., 2012, \apj, 748, 127

\bibitem[{{Foley} {et~al}\mbox{.}(2012{\natexlab{a}}){Foley}, {Challis},
  {Chornock}, {Ganeshalingam}, {Li}, {Marion}, {Morrell}, {Pignata},
  {Stritzinger}, {Silverman}, {Wang}, {Anderson}, {Filippenko}, {Freedman},
  {Hamuy}, {Jha}, {Kirshner}, {McCully}, {Persson}, {Phillips}, {Reichart}, \&
  {Soderberg}}]{Foley12:iax}
{Foley} R.~J. {et~al.}, 2012{\natexlab{a}}, ArXiv e-prints, 1212.2209

\bibitem[{{Foley} {et~al}\mbox{.}(2012{\natexlab{b}}){Foley}, {Challis},
  {Filippenko}, {Ganeshalingam}, {Landsman}, {Li}, {Marion}, {Silverman},
  {Beaton}, {Bennert}, {Cenko}, {Childress}, {Guhathakurta}, {Jiang},
  {Kalirai}, {Kirshner}, {Stockton}, {Tollerud}, {Vink{\'o}}, {Wheeler}, \&
  {Woo}}]{Foley12:09ig}
{Foley} R.~J. {et~al.}, 2012{\natexlab{b}}, \apj, 744, 38

\bibitem[{{Foley} {et~al}\mbox{.}(2012{\natexlab{c}}){Foley}, {Kromer}, {Howie
  Marion}, {Pignata}, {Stritzinger}, {Taubenberger}, {Challis}, {Filippenko},
  {Folatelli}, {Hillebrandt}, {Hsiao}, {Kirshner}, {Li}, {Morrell},
  {R{\"o}pke}, {Ciaraldi-Schoolmann}, {Seitenzahl}, {Silverman}, {Simcoe},
  {Berta}, {Ivarsen}, {Newton}, {Nysewander}, \& {Reichart}}]{Foley12:11iv}
{Foley} R.~J. {et~al.}, 2012{\natexlab{c}}, \apjl, 753, L5

\bibitem[{{Foley} {et~al}\mbox{.}(2012{\natexlab{d}}){Foley}, {Simon}, {Burns},
  {Gal-Yam}, {Hamuy}, {Kirshner}, {Morrell}, {Phillips}, {Shields}, \&
  {Sternberg}}]{Foley12:csm}
{Foley} R.~J. {et~al.}, 2012{\natexlab{d}}, \apj, 752, 101

\bibitem[{{Garavini} {et~al}\mbox{.}(2004){Garavini}, {Folatelli}, {Goobar},
  {Nobili}, {Aldering}, {Amadon}, {Amanullah}, {Astier}, {Balland}, {Blanc},
  {Burns}, {Conley}, {Dahl{\'e}n}, {Deustua}, {Ellis}, {Fabbro}, {Fan}, {Frye},
  {Gates}, {Gibbons}, {Goldhaber}, {Goldman}, {Groom}, {Haissinski}, {Hardin},
  {Hook}, {Howell}, {Kasen}, {Kent}, {Kim}, {Knop}, {Lee}, {Lidman}, {Mendez},
  {Miller}, {Moniez}, {Mour{\~a}o}, {Newberg}, {Nugent}, {Pain}, {Perdereau},
  {Perlmutter}, {Prasad}, {Quimby}, {Raux}, {Regnault}, {Rich}, {Richards},
  {Ruiz-Lapuente}, {Sainton}, {Schaefer}, {Schahmaneche}, {Smith}, {Spadafora},
  {Stanishev}, {Walton}, {Wang}, \& {Wood-Vasey}}]{Garavini04}
{Garavini} G. {et~al.}, 2004, \aj, 128, 387

\bibitem[{{Garavini} {et~al}\mbox{.}(2007){Garavini}, {Nobili}, {Taubenberger},
  {Pastorello}, {Elias-Rosa}, {Stanishev}, {Blanc}, {Benetti}, {Goobar},
  {Mazzali}, {Sanchez}, {Salvo}, {Schmidt}, \& {Hillebrandt}}]{Garavini07:05cf}
{Garavini} G. {et~al.}, 2007, \aap, 471, 527

\bibitem[{{Gerardy} {et~al}\mbox{.}(2004){Gerardy}, {H{\"o}flich}, {Fesen},
  {Marion}, {Nomoto}, {Quimby}, {Schaefer}, {Wang}, \& {Wheeler}}]{Gerardy04}
{Gerardy} C.~L. {et~al.}, 2004, \apj, 607, 391

\bibitem[{{Green} {et~al}\mbox{.}(2012){Green}, {Schechter}, {Baltay}, {Bean},
  {Bennett}, {Brown}, {Conselice}, {Donahue}, {Fan}, {Gaudi}, {Hirata},
  {Kalirai}, {Lauer}, {Nichol}, {Padmanabhan}, {Perlmutter}, {Rauscher},
  {Rhodes}, {Roellig}, {Stern}, {Sumi}, {Tanner}, {Wang}, {Weinberg}, {Wright},
  {Gehrels}, {Sambruna}, {Traub}, {Anderson}, {Cook}, {Garnavich},
  {Hillenbrand}, {Ivezic}, {Kerins}, {Lunine}, {McDonald}, {Penny}, {Phillips},
  {Rieke}, {Riess}, {van der Marel}, {Barry}, {Cheng}, {Content}, {Cutri},
  {Goullioud}, {Grady}, {Helou}, {Jackson}, {Kruk}, {Melton}, {Peddie},
  {Rioux}, \& {Seiffert}}]{Green12}
{Green} J. {et~al.}, 2012, ArXiv e-prints, 1208.4012

\bibitem[{{Hachinger} {et~al}\mbox{.}(2012){Hachinger}, {Mazzali}, {Sullivan},
  {Ellis}, {Maguire}, {Gal-Yam}, {Howell}, {Nugent}, {Baron}, {Cooke},
  {Arcavi}, {Bersier}, {Dilday}, {James}, {Kasliwal}, {Kulkarni}, {Ofek},
  {Laher}, {Parrent}, {Surace}, {Yaron}, \& {Walker}}]{Hachinger12:10jn}
{Hachinger} S. {et~al.}, 2012, ArXiv e-prints, 1208.1267

\bibitem[{{Hatano} {et~al}\mbox{.}(1999){Hatano}, {Branch}, {Fisher}, {Baron},
  \& {Filippenko}}]{Hatano99:94d}
{Hatano} K., {Branch} D., {Fisher} A., {Baron} E., {Filippenko} A.~V., 1999,
  \apj, 525, 881

\bibitem[{{H\"{o}flich}(1995)}]{Hoflich95}
{H\"{o}flich} P., 1995, \apj, 443, 89

\bibitem[{{H\"{o}flich} {et~al}\mbox{.}(1998){H\"{o}flich}, {Wheeler}, \&
  {Thielemann}}]{Hoflich98}
{H\"{o}flich} P., {Wheeler} J.~C., {Thielemann} F.-K., 1998, \apj, 495, 617

\bibitem[{{Hogg} {et~al}\mbox{.}(2010){Hogg}, {Bovy}, \& {Lang}}]{Hogg10}
{Hogg} D.~W., {Bovy} J., {Lang} D., 2010, ArXiv e-prints, 1008.4686

\bibitem[{{Horesh} {et~al}\mbox{.}(2012){Horesh}, {Kulkarni}, {Fox},
  {Carpenter}, {Kasliwal}, {Ofek}, {Quimby}, {Gal-Yam}, {Cenko}, {de Bruyn},
  {Kamble}, {Wijers}, {van der Horst}, {Kouveliotou}, {Podsiadlowski},
  {Sullivan}, {Maguire}, {Howell}, {Nugent}, {Gehrels}, {Law}, {Poznanski}, \&
  {Shara}}]{Horesh12}
{Horesh} A. {et~al.}, 2012, \apj, 746, 21

\bibitem[{{Kasen} {et~al}\mbox{.}(2003){Kasen}, {Nugent}, {Wang}, {Howell},
  {Wheeler}, {H{\"o}flich}, {Baade}, {Baron}, \& {Hauschildt}}]{Kasen03}
{Kasen} D. {et~al.}, 2003, \apj, 593, 788

\bibitem[{{Kasen} \& {Plewa}(2007)}]{Kasen07:asym}
{Kasen} D., {Plewa} T., 2007, \apj, 662, 459

\bibitem[{{Kelly}(2007)}]{Kelly07}
{Kelly} B.~C., 2007, \apj, 665, 1489

\bibitem[{{Kirshner} {et~al}\mbox{.}(1993){Kirshner}, {Jeffery}, {Leibundgut},
  {Challis}, {Sonneborn}, {Phillips}, {Suntzeff}, {Smith}, {Winkler}, {Winge},
  {Hamuy}, {Hunter}, {Roth}, {Blades}, {Branch}, {Chevalier}, {Fransson},
  {Panagia}, {Wagoner}, {Wheeler}, \& {Harkness}}]{Kirshner93}
{Kirshner} R.~P. {et~al.}, 1993, \apj, 415, 589

\bibitem[{{Lentz} {et~al}\mbox{.}(2000){Lentz}, {Baron}, {Branch},
  {Hauschildt}, \& {Nugent}}]{Lentz00}
{Lentz} E.~J., {Baron} E., {Branch} D., {Hauschildt} P.~H., {Nugent} P.~E.,
  2000, \apj, 530, 966

\bibitem[{{Lentz} {et~al}\mbox{.}(2001){Lentz}, {Baron}, {Branch}, \&
  {Hauschildt}}]{Lentz01}
{Lentz} E.~J., {Baron} E., {Branch} D., {Hauschildt} P.~H., 2001, \apj, 557,
  266

\bibitem[{{Li} {et~al}\mbox{.}(2001){Li}, {Filippenko}, {Gates}, {Chornock},
  {Gal-Yam}, {Ofek}, {Leonard}, {Modjaz}, {Rich}, {Riess}, \&
  {Treffers}}]{Li01:00cx}
{Li} W. {et~al.}, 2001, \pasp, 113, 1178

\bibitem[{{Maguire} {et~al}\mbox{.}(2012){Maguire}, {Sullivan}, {Ellis},
  {Nugent}, {Howell}, {Gal-Yam}, {Cooke}, {Mazzali}, {Pan}, {Dilday}, {Thomas},
  {Arcavi}, {Ben-Ami}, {Bersier}, {Bianco}, {Fulton}, {Hook}, {Horesh},
  {Hsiao}, {James}, {Podsiadlowski}, {Walker}, {Yaron}, {Kasliwal}, {Laher},
  {Law}, {Ofek}, {Poznanski}, \& {Surace}}]{Maguire12}
{Maguire} K. {et~al.}, 2012, \mnras, 426, 2359

\bibitem[{{Margutti} {et~al}\mbox{.}(2012){Margutti}, {Soderberg}, {Chomiuk},
  {Chevalier}, {Hurley}, {Milisavljevic}, {Foley}, {Hughes}, {Slane},
  {Fransson}, {Moe}, {Barthelmy}, {Boynton}, {Briggs}, {Connaughton}, {Costa},
  {Cummings}, {Del Monte}, {Enos}, {Fellows}, {Feroci}, {Fukazawa}, {Gehrels},
  {Goldsten}, {Golovin}, {Hanabata}, {Harshman}, {Krimm}, {Litvak},
  {Makishima}, {Marisaldi}, {Mitrofanov}, {Murakami}, {Ohno}, {Palmer},
  {Sanin}, {Starr}, {Svinkin}, {Takahashi}, {Tashiro}, {Terada}, \&
  {Yamaoka}}]{Margutti12}
{Margutti} R. {et~al.}, 2012, \apj, 751, 134

\bibitem[{{Matheson} {et~al}\mbox{.}(2008){Matheson}, {Kirshner}, {Challis},
  {Jha}, {Garnavich}, {Berlind}, {Calkins}, {Blondin}, {Balog}, {Bragg},
  {Caldwell}, {Dendy Concannon}, {Falco}, {Graves}, {Huchra}, {Kuraszkiewicz},
  {Mader}, {Mahdavi}, {Phelps}, {Rines}, {Song}, \& {Wilkes}}]{Matheson08}
{Matheson} T. {et~al.}, 2008, \aj, 135, 1598

\bibitem[{{Matheson} {et~al}\mbox{.}(2012){Matheson}, {Joyce}, {Allen}, {Saha},
  {Silva}, {Wood-Vasey}, {Adams}, {Anderson}, {Beck}, {Bentz}, {Bershady},
  {Binkert}, {Butler}, {Camarata}, {Eigenbrot}, {Everett}, {Gallagher},
  {Garnavich}, {Glikman}, {Harbeck}, {Hargis}, {Herbst}, {Horch}, {Howell},
  {Jha}, {Kaczmarek}, {Knezek}, {Manne-Nicholas}, {Mathieu}, {Meixner},
  {Milliman}, {Power}, {Rajagopal}, {Reetz}, {Rhode}, {Schechtman-Rook},
  {Schwamb}, {Schweiker}, {Simmons}, {Simon}, {Summers}, {Young}, {Weyant},
  {Wilcots}, {Will}, \& {Williams}}]{Matheson12}
{Matheson} T. {et~al.}, 2012, \apj, 754, 19

\bibitem[{{Mazzali} {et~al}\mbox{.}(2005{\natexlab{a}}){Mazzali}, {Benetti},
  {Altavilla}, {Blanc}, {Cappellaro}, {Elias-Rosa}, {Garavini}, {Goobar},
  {Harutyunyan}, {Kotak}, {Leibundgut}, {Lundqvist}, {Mattila}, {Mendez},
  {Nobili}, {Pain}, {Pastorello}, {Patat}, {Pignata}, {Podsiadlowski},
  {Ruiz-Lapuente}, {Salvo}, {Schmidt}, {Sollerman}, {Stanishev}, {Stehle},
  {Tout}, {Turatto}, \& {Hillebrandt}}]{Mazzali05:hvf}
{Mazzali} P.~A. {et~al.}, 2005{\natexlab{a}}, \apjl, 623, L37

\bibitem[{{Mazzali} {et~al}\mbox{.}(2005{\natexlab{b}}){Mazzali}, {Benetti},
  {Stehle}, {Branch}, {Deng}, {Maeda}, {Nomoto}, \& {Hamuy}}]{Mazzali05:99ee}
{Mazzali} P.~A., {Benetti} S., {Stehle} M., {Branch} D., {Deng} J., {Maeda} K.,
  {Nomoto} K., {Hamuy} M., 2005{\natexlab{b}}, \mnras, 357, 200

\bibitem[{{Nugent} {et~al}\mbox{.}(1995){Nugent}, {Phillips}, {Baron},
  {Branch}, \& {Hauschildt}}]{Nugent95}
{Nugent} P., {Phillips} M., {Baron} E., {Branch} D., {Hauschildt} P., 1995,
  \apjl, 455, L147

\bibitem[{{Nugent} {et~al}\mbox{.}(1997){Nugent}, {Baron}, {Branch}, {Fisher},
  \& {Hauschildt}}]{Nugent97}
{Nugent} P., {Baron} E., {Branch} D., {Fisher} A., {Hauschildt} P.~H., 1997,
  \apj, 485, 812

\bibitem[{{Nugent} {et~al}\mbox{.}(2011){Nugent}, {Sullivan}, {Cenko},
  {Thomas}, {Kasen}, {Howell}, {Bersier}, {Bloom}, {Kulkarni}, {Kandrashoff},
  {Filippenko}, {Silverman}, {Marcy}, {Howard}, {Isaacson}, {Maguire},
  {Suzuki}, {Tarlton}, {Pan}, {Bildsten}, {Fulton}, {Parrent}, {Sand},
  {Podsiadlowski}, {Bianco}, {Dilday}, {Graham}, {Lyman}, {James}, {Kasliwal},
  {Law}, {Quimby}, {Hook}, {Walker}, {Mazzali}, {Pian}, {Ofek}, {Gal-Yam}, \&
  {Poznanski}}]{Nugent11}
{Nugent} P.~E. {et~al.}, 2011, \nat, 480, 344

\bibitem[{{Parrent} {et~al}\mbox{.}(2012){Parrent}, {Howell}, {Friesen},
  {Thomas}, {Fesen}, {Milisavljevic}, {Bianco}, {Dilday}, {Nugent}, {Baron},
  {Arcavi}, {Ben-Ami}, {Bersier}, {Bildsten}, {Bloom}, {Cao}, {Cenko},
  {Filippenko}, {Gal-Yam}, {Kasliwal}, {Konidaris}, {Kulkarni}, {Law},
  {Levitan}, {Maguire}, {Mazzali}, {Ofek}, {Pan}, {Polishook}, {Poznanski},
  {Quimby}, {Silverman}, {Sternberg}, {Sullivan}, {Walker}, {Xu}, {Buton}, \&
  {Pereira}}]{Parrent12}
{Parrent} J.~T. {et~al.}, 2012, \apjl, 752, L26

\bibitem[{{Quimby} {et~al}\mbox{.}(2006){Quimby}, {H{\"o}flich}, {Kannappan},
  {Rykoff}, {Rujopakarn}, {Akerlof}, {Gerardy}, \& {Wheeler}}]{Quimby06:05cg}
{Quimby} R., {H{\"o}flich} P., {Kannappan} S.~J., {Rykoff} E., {Rujopakarn} W.,
  {Akerlof} C.~W., {Gerardy} C.~L., {Wheeler} J.~C., 2006, \apj, 636, 400

\bibitem[{{Rodney} {et~al}\mbox{.}(2012){Rodney}, {Riess}, {Dahlen},
  {Strolger}, {Ferguson}, {Hjorth}, {Frederiksen}, {Weiner}, {Mobasher},
  {Casertano}, {Jones}, {Challis}, {Faber}, {Filippenko}, {Garnavich}, {Graur},
  {Grogin}, {Hayden}, {Jha}, {Kirshner}, {Kocevski}, {Koekemoer}, {McCully},
  {Patel}, {Rajan}, \& {Scarlata}}]{Rodney12}
{Rodney} S.~A. {et~al.}, 2012, \apj, 746, 5

\bibitem[{{R{\"o}pke} {et~al}\mbox{.}(2012){R{\"o}pke}, {Kromer}, {Seitenzahl},
  {Pakmor}, {Sim}, {Taubenberger}, {Ciaraldi-Schoolmann}, {Hillebrandt},
  {Aldering}, {Antilogus}, {Baltay}, {Benitez-Herrera}, {Bongard}, {Buton},
  {Canto}, {Cellier-Holzem}, {Childress}, {Chotard}, {Copin}, {Fakhouri},
  {Fink}, {Fouchez}, {Gangler}, {Guy}, {Hachinger}, {Hsiao}, {Chen},
  {Kerschhaggl}, {Kowalski}, {Nugent}, {Paech}, {Pain}, {Pecontal}, {Pereira},
  {Perlmutter}, {Rabinowitz}, {Rigault}, {Runge}, {Saunders}, {Smadja},
  {Suzuki}, {Tao}, {Thomas}, {Tilquin}, \& {Wu}}]{Roepke12}
{R{\"o}pke} F.~K. {et~al.}, 2012, \apjl, 750, L19

\bibitem[{{Shappee} {et~al}\mbox{.}(2013){Shappee}, {Stanek}, {Pogge}, \&
  {Garnavich}}]{Shappee13}
{Shappee} B.~J., {Stanek} K.~Z., {Pogge} R.~W., {Garnavich} P.~M., 2013, \apjl,
  762, L5

\bibitem[{{Silverman} {et~al}\mbox{.}(2012){Silverman}, {Ganeshalingam}, {Li},
  \& {Filippenko}}]{Silverman12:lc}
{Silverman} J.~M., {Ganeshalingam} M., {Li} W., {Filippenko} A.~V., 2012,
  \mnras, 425, 1889

\bibitem[{{Stanishev} {et~al}\mbox{.}(2007){Stanishev}, {Goobar}, {Benetti},
  {Kotak}, {Pignata}, {Navasardyan}, {Mazzali}, {Amanullah}, {Garavini},
  {Nobili}, {Qiu}, {Elias-Rosa}, {Ruiz-Lapuente}, {Mendez}, {Meikle}, {Patat},
  {Pastorello}, {Altavilla}, {Gustafsson}, {Harutyunyan}, {Iijima},
  {Jakobsson}, {Kichizhieva}, {Lundqvist}, {Mattila}, {Melinder}, {Pavlenko},
  {Pavlyuk}, {Sollerman}, {Tsvetkov}, {Turatto}, \&
  {Hillebrandt}}]{Stanishev07:03du}
{Stanishev} V. {et~al.}, 2007, \aap, 469, 645

\bibitem[{{Suzuki} {et~al}\mbox{.}(2012){Suzuki}, {Rubin}, {Lidman},
  {Aldering}, {Amanullah}, {Barbary}, {Barrientos}, {Botyanszki}, {Brodwin},
  {Connolly}, {Dawson}, {Dey}, {Doi}, {Donahue}, {Deustua}, {Eisenhardt},
  {Ellingson}, {Faccioli}, {Fadeyev}, {Fakhouri}, {Fruchter}, {Gilbank},
  {Gladders}, {Goldhaber}, {Gonzalez}, {Goobar}, {Gude}, {Hattori}, {Hoekstra},
  {Hsiao}, {Huang}, {Ihara}, {Jee}, {Johnston}, {Kashikawa}, {Koester},
  {Konishi}, {Kowalski}, {Linder}, {Lubin}, {Melbourne}, {Meyers}, {Morokuma},
  {Munshi}, {Mullis}, {Oda}, {Panagia}, {Perlmutter}, {Postman}, {Pritchard},
  {Rhodes}, {Ripoche}, {Rosati}, {Schlegel}, {Spadafora}, {Stanford},
  {Stanishev}, {Stern}, {Strovink}, {Takanashi}, {Tokita}, {Wagner}, {Wang},
  {Yasuda}, {Yee}, \& {Supernova Cosmology Project}}]{Suzuki12}
{Suzuki} N. {et~al.}, 2012, \apj, 746, 85

\bibitem[{{Tanaka} {et~al}\mbox{.}(2008){Tanaka}, {Mazzali}, {Benetti},
  {Nomoto}, {Elias-Rosa}, {Kotak}, {Pignata}, {Stanishev}, \&
  {Hachinger}}]{Tanaka08}
{Tanaka} M. {et~al.}, 2008, \apj, 677, 448

\bibitem[{{Tanaka} {et~al}\mbox{.}(2011){Tanaka}, {Mazzali}, {Stanishev},
  {Maurer}, {Kerzendorf}, \& {Nomoto}}]{Tanaka11}
{Tanaka} M., {Mazzali} P.~A., {Stanishev} V., {Maurer} I., {Kerzendorf} W.~E.,
  {Nomoto} K., 2011, \mnras, 410, 1725

\bibitem[{{Valenti} {et~al}\mbox{.}(2009){Valenti}, {Pastorello}, {Cappellaro},
  {Benetti}, {Mazzali}, {Manteca}, {Taubenberger}, {Elias-Rosa}, {Ferrando},
  {Harutyunyan}, {Hentunen}, {Nissinen}, {Pian}, {Turatto}, {Zampieri}, \&
  {Smartt}}]{Valenti09}
{Valenti} S. {et~al.}, 2009, \nat, 459, 674

\bibitem[{{Wang} {et~al}\mbox{.}(2003){Wang}, {Baade}, {H{\"o}flich},
  {Khokhlov}, {Wheeler}, {Kasen}, {Nugent}, {Perlmutter}, {Fransson}, \&
  {Lundqvist}}]{Wang03}
{Wang} L. {et~al.}, 2003, \apj, 591, 1110

\bibitem[{{Wang} {et~al}\mbox{.}(2009{\natexlab{a}}){Wang}, {Filippenko},
  {Ganeshalingam}, {Li}, {Silverman}, {Wang}, {Chornock}, {Foley}, {Gates},
  {Macomber}, {Serduke}, {Steele}, \& {Wong}}]{Wang09:2pop}
{Wang} X. {et~al.}, 2009{\natexlab{a}}, \apjl, 699, L139

\bibitem[{{Wang} {et~al}\mbox{.}(2009{\natexlab{b}}){Wang}, {Li}, {Filippenko},
  {Foley}, {Kirshner}, {Modjaz}, {Bloom}, {Brown}, {Carter}, {Friedman},
  {Gal-Yam}, {Ganeshalingam}, {Hicken}, {Krisciunas}, {Milne}, {Silverman},
  {Suntzeff}, {Wood-Vasey}, {Cenko}, {Challis}, {Fox}, {Kirkman}, {Li}, {Li},
  {Malkan}, {Moore}, {Reitzel}, {Rich}, {Serduke}, {Shang}, {Steele}, {Swift},
  {Tao}, {Wong}, \& {Zhang}}]{Wang09:05cf}
{Wang} X. {et~al.}, 2009{\natexlab{b}}, \apj, 697, 380

\bibitem[{{Williams} {et~al}\mbox{.}(2010){Williams}, {Bureau}, \&
  {Cappellari}}]{Williams10}
{Williams} M.~J., {Bureau} M., {Cappellari} M., 2010, \mnras, 409, 1330

\bibitem[{{Yaron} \& {Gal-Yam}(2012)}]{Yaron12}
{Yaron} O., {Gal-Yam} A., 2012, \pasp, 124, 668

\end{thebibliography}

\onecolumn
\begin{deluxetable}{l@{ }l@{ }r@{ }r@{ }r@{ }r@{ }r@{ }r@{ }r}
\tabletypesize{\scriptsize}
\tablewidth{0pt}
\tablecaption{Derived Quantities for \citetalias{Maguire12} Sample\label{t:data}}
\tablehead{
\colhead{SN} &
\colhead{$z$} &
\colhead{Eff.\ Phase} &
\colhead{Stretch\tablenotemark{b}} &
\colhead{\citetalias{Maguire12} \vca} &
\colhead{Phase-corrected \citetalias{Maguire12}} &
\colhead{Blue \vca} &
\colhead{Red \vca} &
\colhead{Ca/Si} \\
\colhead{} &
\colhead{} &
\colhead{(d) \tablenotemark{a,b}} &
\colhead{} &
\colhead{($10^{3}$~\kms)\tablenotemark{b}} &
\colhead{\vca\ ($10^{3}$~\kms)\tablenotemark{b,c}} &
\colhead{($10^{3}$~\kms)\tablenotemark{d}} &
\colhead{($10^{3}$~\kms)\tablenotemark{d}} &
\colhead{Ratio}}

\startdata

PTF09dlc & 0.0666   &   2.8  & 1.05 (0.03) & $-16.99$ (0.17) & $-17.78$ (0.67) & $-19.25$ (0.12) & $-10.47$ (0.19) & 1.709 (0.050) \\
PTF09dnl & 0.019    &   1.3  & 1.05 (0.02) & $-16.81$ (0.15) & $-17.17$ (0.33) & $-18.87$ (0.05) &  $-9.71$ (0.07) & 1.987 (0.047) \\
PTF09fox & 0.0707   &   2.6  & 0.92 (0.04) & $-14.40$ (0.04) & $-15.14$ (0.62) & $-16.99$ (0.33) & $-10.28$ (0.31) & 1.278 (0.041) \\
PTF09foz & 0.05331  &   2.8  & 0.87 (0.06) & $-14.28$ (0.11) & $-15.07$ (0.67) & $-17.32$ (0.30) & $-10.02$ (0.28) & 1.058 (0.006) \\
PTF10bjs & 0.0303   &   1.9  & 1.08 (0.02) & $-16.47$ (0.22) & $-17.02$ (0.50) & $-21.68$ (0.10) & $-13.37$ (0.07) & 0.790 (0.004) \\
PTF10hdv & 0.0542   &   3.3  & 1.05 (0.07) & $-17.03$ (0.18) & $-17.94$ (0.78) & $-19.71$ (0.18) & $-10.94$ (0.22) & 1.637 (0.048) \\
PTF10hmv & 0.033    &   2.5  & 1.15 (0.01) & $-15.02$ (0.11) & $-15.72$ (0.59) & $-17.84$ (0.16) &  $-9.92$ (0.18) & 1.420 (0.026) \\
PTF10icb & 0.0088   &   0.8  & 0.99 (0.03) & $-12.61$ (0.03) & $-12.85$ (0.20) & $-16.99$ (0.06) & $-10.25$ (0.05) & 0.733 (0.004) \\
PTF10mwb & 0.0312   & $-0.4$ & 0.94 (0.03) & $-13.09$ (0.03) & $-12.99$ (0.09) & $-16.99$ (0.12) & $-10.24$ (0.12) & 0.903 (0.008) \\
PTF10qjq & 0.0288   &   3.5  & 0.96 (0.02) & $-11.87$ (0.08) & $-12.86$ (0.83) & $-17.64$ (0.24) & $-10.80$ (0.12) & 0.459 (0.022) \\
PTF10tce & 0.039716 &   3.5  & 1.07 (0.02) & $-16.02$ (0.05) & $-17.00$ (0.81) & $-19.81$ (0.17) & $-12.35$ (0.16) & 0.850 (0.016) \\
PTF10wnm & 0.0645   &   4.1  & 1.01 (0.03) & $-12.90$ (0.07) & $-14.04$ (0.96) & $-17.46$ (0.40) & $-10.66$ (0.24) & 0.579 (0.034) \\
PTF10xyt & 0.0484   &   3.2  & 1.07 (0.04) & $-15.01$ (0.08) & $-15.90$ (0.74) & $-18.74$ (0.48) & $-11.62$ (0.38) & 0.810 (0.026) \\
SN2009le & 0.01703  &   0.3  & 1.08 (0.01) & $-16.08$ (0.12) & $-16.16$ (0.14) & $-19.98$ (0.10) & $-12.24$ (0.10) & 0.966 (0.001)

\enddata

\tablenotetext{a}{Effective phase is the measured phase divided by the
  stretch.}

\tablenotetext{b}{As reported by \citetalias{Maguire12}.}

\tablenotetext{c}{Measured \vca\ corrected by \citetalias{Maguire12}
  velocity gradient of $280 \pm 230$~km~s$^{-1}$~d$^{-1}$.}

\tablenotetext{d}{Assuming a rest wavelength of 3945~\AA.}

\end{deluxetable}

\begin{deluxetable}{lrrrrrrrrrr}
\tabletypesize{\scriptsize}
\tablewidth{0pt}
\tablecaption{{\small SYNOW} Model Parameters for SNe~2010ae and 2011fe\label{t:models}}
\tablehead{
\colhead{Parameter} &
\colhead{\ion{O}{I}} &
\colhead{\ion{Na}{I}} &
\colhead{\ion{Mg}{II}} &
\colhead{\ion{Si}{II}} &
\colhead{\ion{S}{II}} &
\colhead{\ion{Ca}{II}} &
\colhead{HV \ion{Ca}{II}} &
\colhead{\ion{Cr}{II}} &
\colhead{\ion{Fe}{III}} &
\colhead{\ion{Co}{II}}}

\startdata

\multicolumn{11}{c}{SN~2011fe} \\
\tableline
\tableline
\multicolumn{11}{c}{\ion{Si}{II} $\lambda 3858$} \\
\tableline

$\tau$        & 0.2 &     0.8 & 0.5 &   3 & 1.2 &   7 &       0 & 60 & 0.5 & 0.4 \\
$v_{e}$       &   5 &       1 &   2 &   2 & 1.5 & 2.5 & \nodata &  2 & 2.5 &   2 \\
$T_{\rm exc}$ &   8 &       8 &   5 &   6 &  12 &  18 & \nodata & 10 &  10 &  10 \\

\tableline
\multicolumn{11}{c}{HV Ca} \\
\tableline

$\tau$        & 0.2 &       0 & 0.5 &   3 & 1.2 &   4 &     1.7 & 60 & 0.5 & 0.4 \\
$v_{e}$       &   5 & \nodata &   2 &   2 & 1.5 & 3.5 &       2 &  2 & 2.5 &   2 \\
$T_{\rm exc}$ &   8 & \nodata &   5 &  15 &  12 &  18 &      18 & 10 &  10 &  10 \\

\tableline
\tableline
\multicolumn{11}{c}{SN~2010ae} \\
\tableline
\tableline
\multicolumn{11}{c}{\ion{Si}{II} $\lambda 3858$} \\
\tableline

$\tau$        & 0.2 &     0.8 & 0.5 &   1 & 1.2 &   4 &       0 & 60 & 0.5 & 0.4 \\
$v_{e}$       &   5 &       1 &   2 &   2 & 1.5 & 2.5 & \nodata &  2 & 2.5 &   2 \\
$T_{\rm exc}$ &   8 &       8 &   5 &   6 &  12 &  18 & \nodata & 10 &  10 &  10 \\

\tableline
\multicolumn{11}{c}{HV Ca} \\
\tableline

$\tau$        & 0.2 &       0 & 0.5 & 1.5 & 1.2 &   4 &       1 & 60 & 0.5 & 0.4 \\
$v_{e}$       &   5 & \nodata &   2 &   2 & 1.5 &   2 &       2 &  2 & 2.5 &   2 \\
$T_{\rm exc}$ &   8 & \nodata &   5 &  15 &  12 &  18 &      18 & 10 &  10 &  10

\enddata

\end{deluxetable}

\twocolumn

\label{lastpage}


\end{document}